\documentclass[twocolumn,prd,superscriptaddress,preprintnumbers,nofootinbib]{revtex4-2}

\usepackage{comment} 

\usepackage{graphicx}
\usepackage{amsmath,bm,amssymb,amsfonts,dsfont}
\usepackage[dvipsnames]{xcolor}
\usepackage[normalem]{ulem}
\usepackage{url}
\usepackage{array}
\usepackage{booktabs}
\usepackage{multirow}
\usepackage{float}
\usepackage[colorlinks  = true,
            linkcolor   = NavyBlue,
            urlcolor    = NavyBlue,
            citecolor   = NavyBlue,
            anchorcolor = NavyBlue]{hyperref}
\usepackage{cprotect}

\usepackage{appendix}

\usepackage[switch]{lineno}

\usepackage{tikz-feynman}
\tikzfeynmanset{compat=1.1.0}

\pdfoutput=1 % if your are submitting a pdflatex (i.e. if you have
\newcommand{\lsim}{\mathrel{\mathop{\kern 0pt \rlap
  {\raise.2ex\hbox{$<$}}}
  \lower.9ex\hbox{\kern-.190em $\sim$}}}
\newcommand{\gsim}{\mathrel{\mathop{\kern 0pt \rlap
  {\raise.2ex\hbox{$>$}}}
  \lower.9ex\hbox{\kern-.190em $\sim$}}}

\def \aap  {Astronomy \& Astrophysics}

\tikzfeynmanset{warn luatex=false} 

\begin{document}

%\preprint{CP3-XXX}
%\preprint{CTPU-PTC-24-31, CERN-TH-2024-164}

%\title{A $U(1)_{L_{i}-L_{j}}$ solution to the {\sc Fermi-Lat} Galactic Center Excess}

\title{Constraining dark matter using 20-year INTEGRAL/IBIS observations I: \\ Primordial black holes}

\author{Jordan Koechler}
\email{jordan.koechler@gmail.com}
\affiliation{Istituto Nazionale di Fisica Nucleare, Sezione di Torino, Via P. Giuria 1, 10125 Torino, Italy}

\author{Pedro De la Torre Luque}
\email{pedro.delatorre@uam.es}
\affiliation{Departamento de F\'isica Te\'orica, M-15, Universidad Aut\'onoma de Madrid, E-28049 Madrid, Spain}
\affiliation{Instituto de F\'isica Te\'orica UAM-CSIC, Universidad Aut\'onoma de Madrid, C/ Nicol\'as Cabrera, 13-15, 28049 Madrid, Spain}

\begin{abstract}
Primordial black holes (PBHs) in the asteroid-mass window remain a viable dark matter candidate. In the lower part of this mass range, Hawking evaporation produces electrons and positrons that generate diffuse hard X-ray emission through inverse Compton (IC) scattering on Galactic radiation fields. We present the first search for this signal using a combined spectral and morphological analysis of the 20-year INTEGRAL/IBIS observations.
We construct a physically motivated model of the Galactic hard X-ray background, including IC emission from cosmic-ray electrons and unresolved accreting white dwarfs, and consistently incorporate the IC contribution from PBH evaporation. By exploiting both spatial and spectral information, we improve the separation between a possible PBH signal and astrophysical backgrounds compared to analyses based only on spectral data.
We derive constraints on the PBH dark matter fraction over a broad mass range, including extended PBH mass functions and rotating PBHs. We assess the impact of uncertainties in cosmic-ray propagation and the Galactic dark matter density profile, finding that the propagation of low-energy electrons and positrons represents the main source of uncertainty in the predicted signal. Despite these limitations, our results provide competitive constraints over a wide range of PBH masses, highlighting the potential of hard X-ray observations as a complementary probe of evaporating PBHs.

\end{abstract}

%\keywords{Dark Matter, Indirect Detection experiments, Monte Carlo event generators.}
%{\let\thefootnote\relax
%\footnotetext{$^\dagger$\,Contact authors} }

\maketitle

\flushbottom

\section{Introduction}

Primordial black holes (PBHs), formed in the early Universe from the collapse of primordial density fluctuations, remain a well-motivated dark matter (DM) candidate~\cite{Carr2021PBHReview, Green2021PBHReview, Zeldovich1967PBH}. While a large fraction of the PBH mass spectrum has now been excluded by a variety of astrophysical and cosmological observations~\cite{Carr2010, Carr2017, Niikura2019, Carr2021PBHReview, Green2021PBHReview}, the asteroid-mass window around $10^{15}$--$10^{21}~\mathrm{g}$ continues to be one of the few regions where PBHs could account for essentially all of the observed DM abundance. PBHs in the lower part of this mass range emit Standard Model particles through Hawking radiation~\cite{Hawking1974BlackHole, MacGibbon1990, Carr2010}, producing photons together with energetic electrons and positrons that can generate detectable diffuse emission across the X-ray and $\gamma$-ray sky. As a result, observations of Galactic diffuse emission provide one of the most promising avenues for probing this remaining viable PBH parameter space~\cite{Cline1997, DeRocco2019, Laha:2019ssq, Calore2021}.

Among these signatures, X-ray observations provide a particularly sensitive probe of PBHs in the mass range where Hawking evaporation produces energetic electrons and positrons. Besides the prompt photon emission directly generated by the evaporation process, these charged particles propagate through the Galactic environment and upscatter ambient radiation fields via inverse Compton (IC) scattering, producing an extended and diffuse X-ray emission. Recent works~\citep{Cirelli:2020bpc,Cirelli:2023tnx,DelaTorreLuque:2023olp,Balaji:2025afr} have demonstrated that this secondary IC component can dominate the observable signal over a significant range of PBH masses and substantially strengthen existing constraints. As in analogous studies of sub-GeV particle DM, the combination of efficient IC production and the excellent sensitivity of modern hard X-ray observations makes this channel especially promising.

The sensitivity of such searches, however, is ultimately limited by our understanding of the diffuse X-ray background. In the hard X-ray band, Galactic emission is expected to arise primarily from IC scattering of cosmic-ray (CR) electrons on the interstellar radiation fields (ISRFs), together with unresolved populations of compact astrophysical sources, including accreting white dwarfs and other stellar remnants. Although considerable progress has been made in modeling these components, significant uncertainties remain in both their spectral and spatial properties. Since the expected PBH signal is itself diffuse and spatially correlated with the Galactic DM distribution, an accurate characterization of the astrophysical backgrounds is essential for obtaining robust constraints.

The Imager on Board the INTEGRAL Satellite (IBIS), onboard the recently decommissioned INTEGRAL observatory, provides an ideal dataset for this purpose. Owing to its coded-mask imaging technique, IBIS has measured not only the spectrum but also the morphology of the diffuse hard X-ray emission across the Galaxy. Recent analyses have exploited these capabilities to extract detailed spatial and spectral information on the diffuse emission~\cite{Krivonos:2024bih, Krivonos:2006px}, enabling significantly improved studies of Galactic high-energy processes and opening new opportunities for indirect searches for DM and evaporating PBHs.

In this work, we exploit the full constraining power of the IBIS observations by performing, for the first time, a joint analysis of their spectral and morphological information to search for the diffuse IC emission generated by electrons and positrons emitted in PBH evaporation. We construct a physically motivated model for the Galactic X-ray background, combining IC emission from CR electrons with unresolved thermal emission, and consistently include the additional IC contribution expected from Hawking radiation. By simultaneously fitting the spatial and spectral characteristics of the data, we are able to disentangle the PBH signal from the dominant astrophysical backgrounds far more effectively than with spectral information alone. This approach substantially improves existing constraints on the fraction of DM that can be composed of PBHs over the mass range where secondary IC emission dominates, while simultaneously providing a more robust characterization of the diffuse hard X-ray sky.

\section{20-year INTEGRAL/IBIS observations}
\label{sec:data}

In this work, we use all publicly available INTEGRAL data collected with the IBIS telescope~\cite{Ubertini:2003ih}, specifically its ISGRI detector~\cite{Lebrun:2003aa}, over the period from May 2003 to February 2024, corresponding to INTEGRAL revolutions 70--2740. The IBIS/ISGRI instrument provides a large field of view, approximately $29^\circ\times29^\circ$, with a fully coded aperture of $9^\circ\times9^\circ$, together with an angular resolution of about $12'$ FWHM. The latter enables the identification and subtraction of point sources, while the former allows for the study of diffuse Galactic emission over wide angular regions.

When the instrumental background is accurately modeled, IBIS/ISGRI can be used as collimated instrument to measure large-scale Galactic emissions after the effective subtraction of point-source contributions~\cite{Krivonos:2006px}. We rely on the publicly available \texttt{ridge} repository associated with Ref.~\cite{Krivonos:2024bih}, which provide the cleaned ISGRI detector count rates, the instrumental-background modeling procedure, and the extraction of morphological profiles and energy spectra of the Galactic hard X-ray and soft-$\gamma$-ray background.

\subsection{Data reduction}

The \texttt{ridge} package defines three broad energy bands for the extraction of Galactic longitude profiles within $|b|<10^\circ$. The first band, $25$--$60$~keV, is dominated by the hard X-ray component associated with the Galactic ridge X-ray emission (GRXE), and its morphology is well traced by near-infrared (NIR) observations~\cite{Krivonos:2006px, 2006A&A...452..169R}. The lower bound of $25$~keV is chosen to mitigate the reduced ISGRI sensitivity at low energies caused by detector degradation over the mission lifetime. The second band, $60$--$80$~keV, corresponds to the transition region between the hard X-ray GRXE component and the emerging soft $\gamma$-ray component. The third band, $80$--$200$~keV, is dominated by the Galactic soft $\gamma$-ray background, mainly generated by CR electrons.

We also extract energy spectra in three extended sky regions. In our analysis, these are defined as follows: (i) The Galactic Bulge region, \texttt{GB20}, a $20^\circ\times20^\circ$ region centered on the Galactic center, $(\ell,b)=(0^\circ,0^\circ)$; (ii) the high-longitude positive window, \texttt{L+80}, a $30^\circ\times20^\circ$ region centered at $(\ell,b)=(80^\circ,0^\circ)$; and (iii) the high-longitude negative window, \texttt{L-80}, a $30^\circ\times20^\circ$ region centered at $(\ell,b)=(-80^\circ,0^\circ)$. The spectra are extracted in 21 predefined logarithmically spaced energy bins between $25$ and $185$~keV.

All three regions span $20^\circ$ in Galactic latitude, matching the latitude range used for the longitudinal profiles, $|b|<10^\circ$. Their longitude ranges are chosen so that the total flux in each region is obtained from the ensemble of observations whose pointing directions fall within the corresponding sky region.

The procedure used by \texttt{ridge} to extract the longitudinal profiles and energy spectra is described in detail in Ref.~\cite{Krivonos:2024bih}. We summarize the main steps here for completeness. The repository contains ISGRI detector count rates, cleaned of point-source contributions, for individual INTEGRAL observations, hereafter science windows (ScWs), in the relevant energy bands and for INTEGRAL revolutions 70--2740.

Since IBIS/ISGRI is used in this analysis as a collimated instrument, the detector count rate associated with a given ScW depends on the angular response of the instrument around the pointing direction. This dependence is encoded in the IBIS/ISGRI collimator response function. The response is approximately piecewise linear, with a break around $4.5^\circ$, corresponding to the transition between the fully coded and partially coded field of view. Within the fully coded region, the response and flux calibration are constrained using observations of the Crab Nebula. In practice, Crab observations with angular separations smaller than $4.5^\circ$ from the ScW pointing direction are used to calibrate the observed Crab count rate and its dependence on off-axis angle.

The ISGRI detector sensitivity evolved significantly over the INTEGRAL mission. The observed Crab count rate therefore changes with revolution number, with the size of the effect depending on the energy band. To account for this time dependence, the Crab count rate is modeled as a smooth function of revolution number. This calibration is then used to convert residual ISGRI detector count rates into Crab flux units for each ScW.

The instrumental background is estimated using high-latitude observations, where the Galactic contribution is expected to be negligible. The model accounts for variations of the detector background within individual INTEGRAL revolutions, using either a linear dependence on orbital phase or an interpolated/constant model when the available high-latitude statistics are insufficient. The approximately isotropic cosmic X-ray background is treated as a constant contribution to this background model.

For each ScW, the residual count rate is obtained by subtracting the predicted background model from the point-source-cleaned detector count rate. This residual is then converted into mCrab units by dividing by the Crab count-rate model evaluated at the corresponding revolution number.

Finally, the longitudinal profiles and energy spectra are obtained by selecting the relevant ScWs for each longitude bin, sky region, and energy band. The distribution of residual fluxes in each selection is then combined using Gaussian-fitting procedure, yielding the flux per IBIS field of view and its associated uncertainty. Each of the three longitude profiles and three energy spectra contains 21 bins, resulting in 126 flux measurements.

\subsection{Treatment of correlations}

The longitudinal bins are not necessarily statistically independent. In particular, the same INTEGRAL revolutions can contribute to several bins, while common data-reduction and background-modeling effects may introduce additional correlations. Because the joint covariance matrix is not provided with the \texttt{ridge} data products, we estimate it using a non-parametric cluster bootstrap in which the INTEGRAL revolution is taken as the resampling unit.

For each bootstrap realization, $N_{\mathrm{rev}}$ revolutions are drawn with replacement from the $N_{\mathrm{rev}}$ contributing revolutions. The same set of resampled revolution multiplicities is applied simultaneously to all 126 measurements, and all ScWs belonging to a selected revolution are kept together. The complete \texttt{ridge} flux-extraction procedure is then repeated for every spatial and spectral bin, producing a bootstrap realization of the full data vector. From $N_b$ of such realizations, the empirical covariance is estimated as
\begin{equation}
    C_{ij}=\frac{1}{N_b-1}\sum_{b=1}^{N_b}\left(F_i^{b}-\overline{F}_i\right)\left(F_j^{b}-\overline{F}_j\right)\;,
\end{equation}
where $F_i^{b}$ is the flux extracted in the bin $i$ at the bootstrap realization $b$, and $\overline{F}_i=\frac{1}{N_b}\sum_{b=1}^{N_b} F_i^{b}$ is its average over the bootstrap realizations.

To reduce sampling noise and improve the numerical conditioning of the covariance matrix, the bootstrap realizations are standardized and Ledoit–Wolf shrinkage~\cite{LEDOIT2004365} is applied to their covariance matrix. The regularized matrix is subsequently rescaled to the original flux units. Its positive definiteness is verified through a Cholesky decomposition before it is used in our analysis. The number of bootstrap realizations $N_b$ is increased until the inferred covariance structure and the resulting parameter constraints are numerically stable.

\section{Hard X-rays and soft $\gamma$-rays from PBH evaporation}
\label{sec:evap}

In this section, we discuss how hard X-rays and soft $\gamma$-rays can be generated from PBH evaporation. Black holes are known to have a finite temperature $T$ which depends solely on their mass $M$. In the case of a Kerr black hole of spin parameter $a=J/M$ (where $J$ is its angular momentum), the black hole temperature is written as follows (with $\hbar = c = k_B=G=1$)
\begin{equation}
    T = \frac{1}{2\pi}\left(\frac{r_+-M}{r_+^2+a^2}\right)\;,
\end{equation}
where $r_+ = M + \sqrt{M^2-a^2}$ is its horizon radius. As the black hole radiate, the emission spectrum of particles is written (with $a^\star=a/M<1$):
\begin{equation}
    \frac{d^2N_i}{dE_idt}=\frac{1}{2\pi}\sum_{\rm d.o.f.}\frac{\Gamma_i(E'_i,M,a^\star)}{e^{E'_i/T}\pm1}
\end{equation}
where the difference with the black body spectrum is encoded in the so-called greybody factor $\Gamma_i$, and the $\pm$ signs depend on the nature of the radiated particles $i$: $+$ for fermions and $-$ for bosons. The sum is performed over all of the degrees of freedom (d.o.f.) of the emitted particles. Finally $E'_i = E_i - m\Omega$ is the energy of the emitted particle where $\Omega=a^\star/(2r_+)$ is the black hole horizon rotation velocity and $m = \{-l,...,l\}$ the projection on the black hole axis of the particle angular momentum $l$. Rotating (Kerr) PBHs exhibit enhanced Hawking emission due angular momentum transfer from the black-hole to the particles (even though their temperature is lower than Schwarschild ones of the same mass), leading to larger fluxes of photons and charged particles compared to their non-rotating counterparts. Consequently, the assumptions regarding the PBH mass and spin distributions have an important impact on the inferred limits on the PBH abundance. Therefore, we consider not only the conventional Schwarzschild case but also investigate the impact of PBH spin on the X-ray signal generated by IC scattering of Hawking-emitted electrons and positrons. This allows us to assess the robustness of our constraints under more general and physically motivated assumptions about the PBH population.
In particular, we explore two extreme values of the spin parameter $a^\star$, which are 0 (Schwarzschild) and 0.9999 (near-extremal).

Most of the emitted particles are in principle unstable, and can decay, hadronise or emit soft radiations. In order to compute the final emission spectra of $\gamma$ and $e^\pm$ from PBH evaporation, we use the numerical code \texttt{BlackHawk} (version 2.2)~\cite{Arbey:2019mbc,Arbey:2021mbl}, which provide the summed spectra of $e^+ + e^-$. Among the hadronization schemes already available in the code, we use \texttt{Hazma}~\cite{Coogan:2019qpu}, since it covers the most relevant energy range for our study.

From the emission spectra of photons and $e^\pm$, we can compute the associated flux of X-rays and soft $\gamma$-rays from the evaporation of Galactic PBHs. For instance, the flux of directly evaporated photons is written, assuming that PBHs makes a fraction $f_{\rm PBH}$ of the entirety of DM:
\begin{equation}
    \frac{d\Phi_\gamma^{\rm direct}}{dE_\gamma d\Omega}=\frac{1}{4\pi}f_{\rm PBH} \mathcal{D}(b,\ell)\int_{M_{\rm min}}^{\infty}\frac{dM}{M}\frac{dN_{\rm PBH}}{dM}\frac{d^2N_\gamma}{dE_\gamma dt}\;,
    \label{eq:directgamma}
\end{equation}
with $\mathcal{D}(b,\ell) = \int_{\rm l.o.s.} ds\,\rho_{\rm DM}(r(s,b,\ell))$ is the so-called $D$-factors, corresponding to line-of-sight (l.o.s.) integral of the Galactic DM density $\rho_{\rm DM}$. In this paper, we adopt the NFW~\cite{Navarro:1995iw} profile as our fiducial case. We also study how the results differ when a cuspier profile than NFW ($\gamma = 1.5$) and the isothermal (cored, $\gamma=0$) profiles~\cite{Cirelli:2010xx, Cirelli:2025rky} in order to evaluate the systematic uncertainties from the choice of the DM profile, since it is not precisely known. 

Although many existing constraints are derived assuming a monochromatic population of non-rotating (Schwarzschild) PBHs, realistic formation scenarios generally predict both extended mass distributions and, potentially, non-zero angular momentum. The PBH mass function depends sensitively on the underlying production mechanism and may span several orders of magnitude rather than being sharply peaked at a single mass. Since the Hawking temperature and evaporation rate depend strongly on the PBH mass, an extended distribution modifies both the spectral shape and overall normalization of the emitted radiation, thereby affecting the interpretation of observational constraints. 
In Eq.~\ref{eq:directgamma}, $dN_{\rm PBH}/dM$ designates the present-day mass distribution of PBHs (normalized to one). Different PBH formation scenarios give rise to different mass distributions.  In this work, we adopt a log-normal distribution with different standard deviation $\sigma$ values
\begin{equation}
    \frac{dN_{\rm PBH}}{dM} = \frac{1}{\sqrt{2\pi \sigma}M}\exp\left(-\frac{\log^2(M/M_{\rm PBH})}{2\sigma^2}\right)\;,
    \label{Eq:lognorm}
\end{equation}
exploring values of $\sigma$ from 0 to 2, with $\sigma \to 0$ corresponding to the monochromatic mass distribution, \emph{i.e.}~all PBHs have the same mass ($dN_{\rm PBH}/dM = \delta(M-M_{\rm PBH})$). Finally, $M_{\rm min}\approx7.5\times10^{14}$ g designates the minimal mass of PBHs today. 

The other source of hard X-rays and soft $\gamma$-rays from PBH evaporation comes from the up-scattering of Galactic ambient photons from evaporated $e^\pm$ through ICS. The flux of photons from IC scattering is written
\begin{equation}
    \label{eq:flux}
    \frac{d\Phi_{\gamma}^{\rm IC}}{dE_\gamma d\Omega}=\frac{1}{4\pi}\frac{1}{E_\gamma}\int_{\rm l.o.s.} ds\;j_{\rm IC}(E_\gamma,\vec{x}(s,b,\ell)\;,
\end{equation}
where $j_{\rm IC}$ is the IC photon emissivity, which is given as the convolution of the IC radiating power $\mathcal{P}_{\rm IC}$~\cite{Cirelli:2009vg} with the DM-produced and propagated $e^\pm$ density $dn_e/dE_e$ at a position $\vec{x}$ of the Milky-Way
\begin{equation}
    \label{eq:ICemissivity}
    j_{\rm IC}(E_\gamma,\vec{x})=\int dE_e\;\mathcal{P}_{\rm IC}(E_\gamma,E_e,\vec{x})\frac{dn_e}{dE_e}(E_e,\vec{x})\;.
\end{equation}
The first step is to compute the spatial distribution and energy spectrum of the propagated DM-produced $e^\pm$ $f_e=dn_e/dE_e$, which is the solution of the diffusion-advection-loss equation~\cite{Ginz&Syr,DRAGON2-1}
\begin{multline}
    \label{eq:prop}
    \frac{\partial}{\partial p_e}\left[p_e^2D_{pp}\frac{\partial}{\partial p_e}\left(\frac{f_e}{p_e^2}\right)+\frac{p_e}{3}\left(\vec{\nabla}\cdot\vec{v}_c\right)f_e-\dot{p}_ef_e\right]+\\+\vec{\nabla}\cdot\left(D\vec{\nabla}f_e-\vec{v}_cf_e\right)+Q_e=0
\end{multline}
where $-\dot{p}_e$ quantifies the momentum losses due to interactions between the DM-produced $e^\pm$ and the interstellar medium (ISM), $\vec{v}_c$ is the Galactic wind velocity responsible for $e^\pm$ convection. $D$ and $D_{pp}$ are, respectively, the spatial and momentum-space diffusion coefficients. $D$ is parameterized as a broken power-law of rigidity $R=|\vec{p}|/q$ with a break at $R_b\simeq 312$ GV~\cite{Genolini:2017dfb}, while $D_{pp}$ characterizes the stochastic reacceleration of $e^\pm$ due to resonant interactions with the turbulent component of the Galactic magnetic fields~\cite{1984acr..book.....B}
\begin{gather}
    D(R)=D_0\beta^\eta\frac{(R/R_0)^\delta}{\left[1+(R/R_b)^{\Delta\delta/s}\right]}\;, \\
    D_{pp}=\frac{4}{3}\frac{1}{\delta(4-\delta^2)(4-\delta)}\frac{v_A^2p^2}{D}\;,
\end{gather}
where $D_0,\eta,R_0,\delta,\Delta\delta,s,v_A$, as well as the halo height $H$, are parameters that are fitted to AMS-02 data, in particular for B, Be and Li ratios to C and O. The values of these parameters are reported in Refs.~\cite{Luque:2021nxb,DelaTorreLuque:2023olp}. In particular, they find a value of the Alfvén velocity $v_A$ (responsible for the stochastic reacceleration of $e^\pm$) of  13.4 km/s, which significantly impacts the distribution of DM-produced $e^\pm$ for DM masses below a few tens of MeV~\cite{DelaTorreLuque:2023olp}. As we will see below, this parameter represents the main uncertainty in our predictions.

Finally, the last key quantity of the propagation equation Eq.~\ref{eq:prop} is the source term $Q_e$, which encodes the injection of PBH-produced $e^\pm$ in the ISM
\begin{equation}
    Q_e(r,E_e)=f_{\rm PBH} \rho_{\rm DM}(r)\int_{M_{\rm min}}^{\infty}\frac{dM}{M}\frac{dN_{\rm PBH}}{dM}\frac{d^2N_e}{dE_e dt}\;,
\end{equation}

To solve Eq.~\ref{eq:prop}, we use the numerical solver \texttt{DRAGON2}~\cite{DRAGON2-1,DRAGON2-2}, and after obtaining the propagated DM-produced $e^\pm$ density $dn_e/dE_e$, we use the integrator code \texttt{HERMES}~\cite{Dundovic:2021ryb} to perform the computation of the IC emissivity (Eq.~\ref{eq:ICemissivity}) and the integration along the l.o.s.~(Eq.~\ref{eq:flux}) in order to obtain the flux of photons from IC $d\Phi_{\gamma}^{\rm IC}/(dE_\gamma d\Omega)$. The output of \texttt{HERMES} is a HEALPix~\cite{Gorski:2004by} map of the Galaxy, where each pixel contains spectra information of the IC flux.

We note that Bremsstrahlung emission is also produced from the interaction of the same electrons with interstellar gas. However, since this contribution scales proportional to the energy of the electrons, it only becomes relevant for $\gamma$-rays above the MeV. We have checked that this contribution remains always below $10\%$ of the IC contribution, and therefore we conservatively neglect it.

\begin{figure*}
    \centering
    \includegraphics[width=\linewidth]{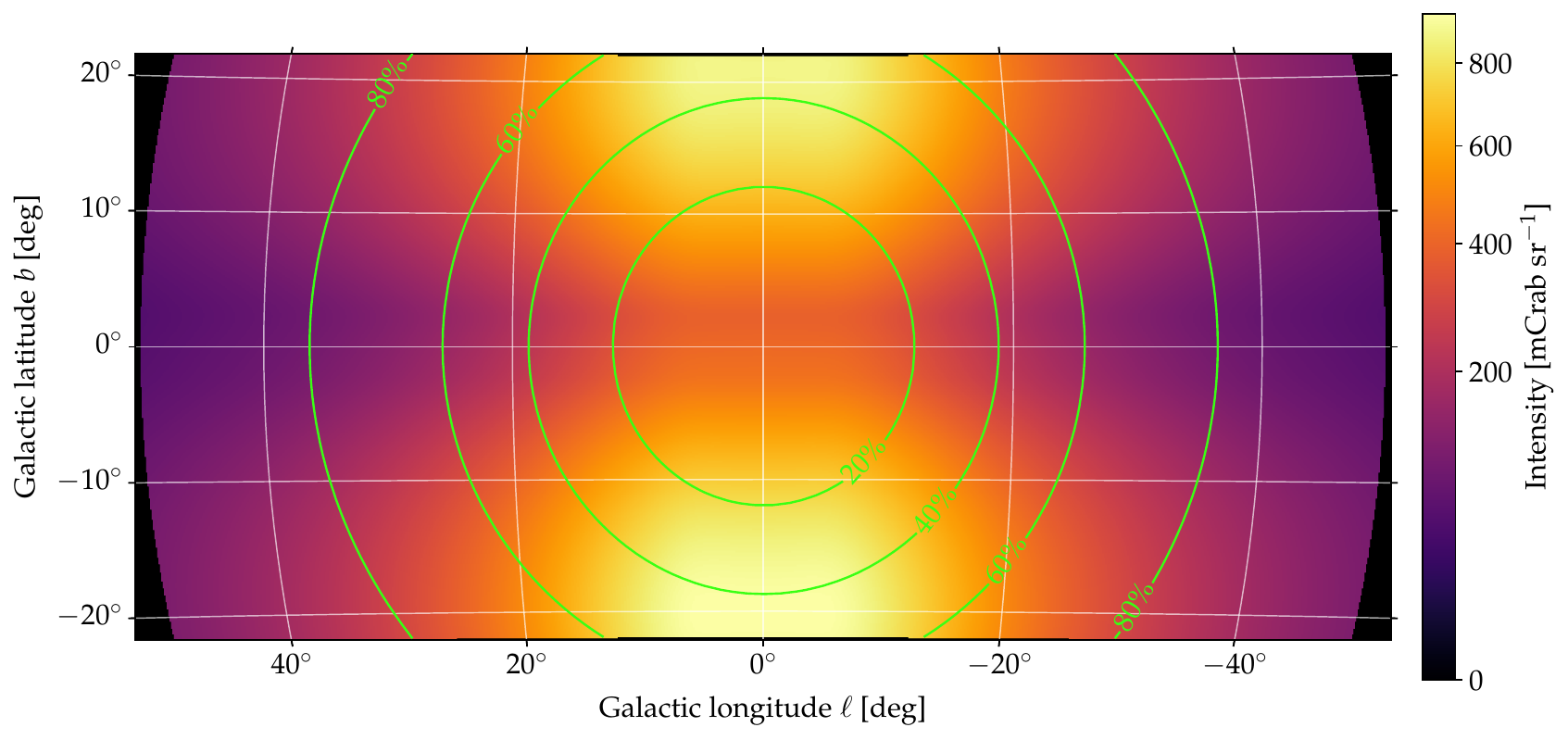}
    \includegraphics[width=\linewidth]{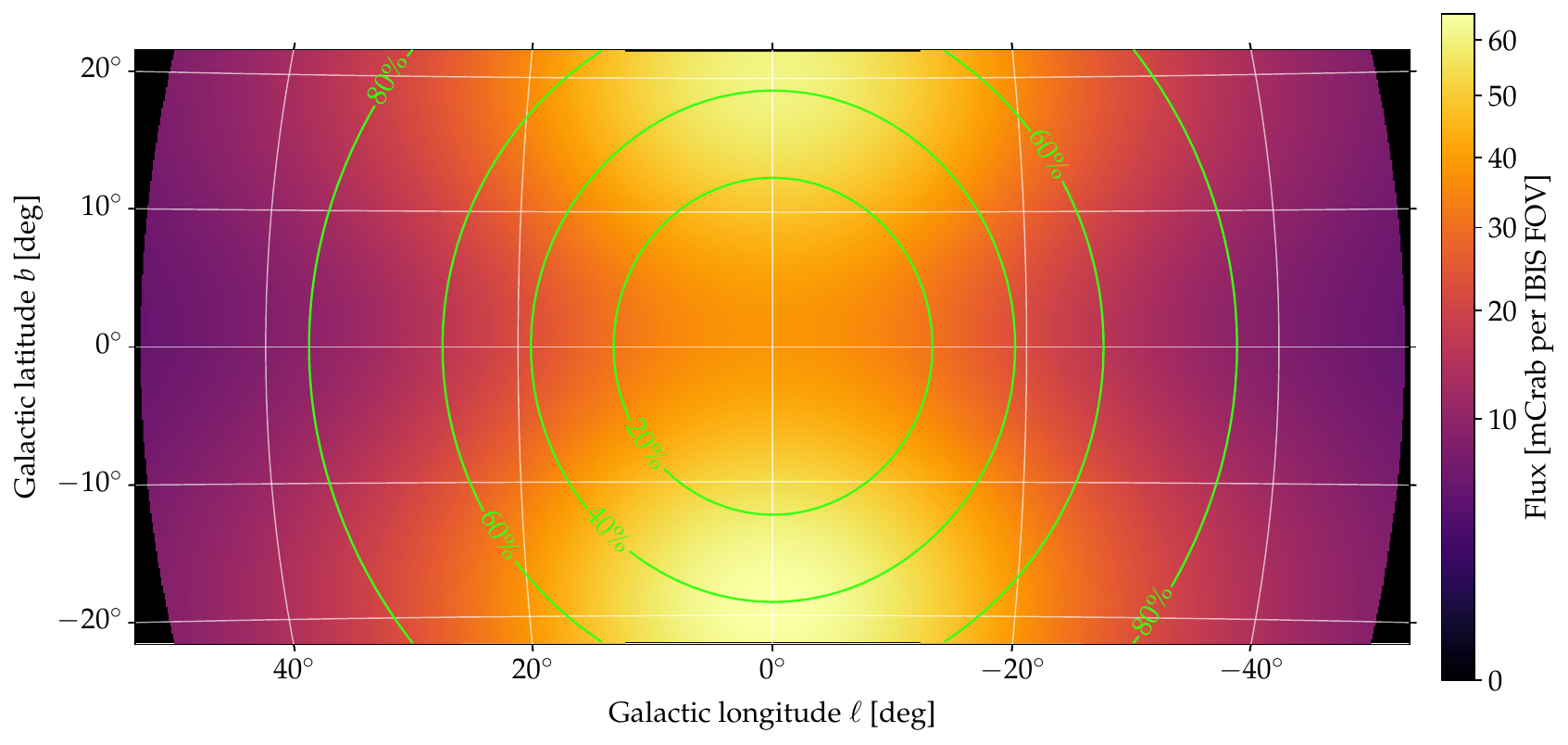}
    \caption{Total emission intensity (FSR + ICS) in the 25--60 keV energy band. The green contours indicate the enclosed fraction of the total FSR flux. The top panel shows the intrinsic emission map, while the bottom panel shows the map convolved with the instrument response function. The maps correspond to $M_{\rm PBH}=10^{16}$ g and $f_{\rm PBH}=10^{-3}$.}
    \label{fig:PBHmap}
\end{figure*}

\section{Astrophysical backgrounds}
Sub-MeV radiation from CR electrons and X-ray emission from accreting white dwarfs constitute the dominant backgrounds in the X-ray sky up to MeV energies. The former arises from IC scattering of CR electrons on ambient radiation fields and dominates above a few tens of keV, while the latter, associated with thermal emission from unresolved accreting systems (related, predominantly, to white dwarfs), is more relevant at lower energies. In the following sections, we detail our modeling of both components.

\subsection{X-ray radiations from CR electrons}
\label{subsec:crel}

CR electrons propagating through the Galaxy are a major source of diffuse X-ray emission via interactions with the interstellar environment. These electrons scatter off the ISRFs, including optical, ultraviolet, infrared, and microwave photons, producing X-rays through IC processes~\cite{Vernetto2016prd}. In addition, their interactions with the interstellar gas generate Bremsstrahlung emission, contributing mostly at the $\gamma$-ray bands. Together, these mechanisms dominate the Galactic emission below a few hundreds of MeV and form the backbone of the diffuse X-ray background which is dominated by IC emission.

To model this emission, we adopt the CR electron distribution in the ISM as developed in Refs.~\cite{delaTorreLuque:2022vhm, DelaTorreLuque:2023zyd}, which is optimized to reproduce both local measurements of electrons and positrons at Earth and the $\gamma$-ray emissivity, observed by Fermi-LAT, down to tens of MeV~\cite{Casandjian:2015hja, delaTorreLuque:2022vhm}. The model accounts for spatial diffusion, energy losses via synchrotron and IC interactions, Coulomb and ionization processes, as well as reacceleration effects. Bremsstrahlung emission from interactions with atomic and molecular gas is computed using detailed spiral-arm gas maps, while IC emission is evaluated using the the ISRF models derived in Ref.~\cite{2016PhRvD..94f3009V}, ensuring a realistic energy-dependent photon output across the X-ray band. Notably, the IC emission dominates at lower energies (keV scale) due to upscattering of CMB photons, whereas Bremsstrahlung contributes more significantly above the MeV, reflecting the energy distribution of CR electrons in the Galaxy.
We note that this model reproduces satisfactorily  a wide range of X-ray and $\gamma$-ray observations (see e.g. Refs.~\cite{laTorreLuquePedro:2024est, DelaTorreLuque:2024zsr}), including measurements at keV energies, which makes it a consistent background choice for our study. %This approach allows the IC spectrum to closely follow the observed X-ray background, capturing both the spectral shape and normalization. 

This model provides predictions for both the spatial distribution and the spectral shape of the X-ray background over a wide energy range, from keV to GeV. However, given the significant uncertainties — primarily associated with the energy density and spatial distribution of the ISRFs, as well as the Galactic electron distribution — we leave the overall normalization of the model free in our fits to the data. As we show, this model naturally reproduces the X-ray observations by IBIS above $\sim 60$~keV across multiple regions of interest, while remains below subdominant at lower energies, where unresolved sources dominate the X-ray background.

\subsection{X-ray emission from accreting white dwarfs}
\label{subsec:WDs}

Accreting white dwarfs (WDs) are a diverse class of systems in which a WD gains mass from a companion star or circumstellar material, leading to a range of observable high‑energy phenomena. In close binary systems such as cataclysmic variables, material transferred via Roche‑lobe overflow forms an accretion disk or column, releasing gravitational potential energy that heats gas to X‑ray emitting temperatures. Depending on the magnetic field of the WD and the accretion regime, the X‑ray spectrum can include optically thin thermal emission from shock‑heated plasma in accretion columns, boundary layers, or disk coronae, as well as softer emission linked to steady nuclear burning on the WD surface in supersoft sources. These mechanisms span a broad X‑ray band, with harder emission typically arising in magnetic systems such as polars and intermediate polars, and softer components in systems with steady or unstable burning processes \cite{Mukai2017, deMartino2020}.

Unresolved populations of accreting WDs are recognized as important contributors to the Galactic X‑ray background, particularly below $\sim 60$ keV. Morphological and spectral studies of the GRXE show that its low‑energy component closely traces the stellar mass distribution of the Milky Way, and its spectral shape is consistent with the integrated emission from large numbers of faint accreting WDs, especially intermediate polars with typical WD masses of order $\sim0.7\,M_\odot$ \cite{Yuasa2012, Krivonos:2006px, Krivonos:2024bih}. In this framework, the cumulative X‑ray output of an unresolved WD population produces a hard continuum that is distinct from truly diffuse processes, with thermal spectra shaped by the physics of accretion flows and shock heating in WD systems.

Modelling the contribution of accreting WDs to the X‑ray background requires population synthesis that accounts for the space density, accretion rates, luminosity functions, and spectral characteristics of the various sub‑classes of systems. Population synthesis studies predict the soft and hard X‑ray luminosities from these populations as a function of stellar age and binary evolution history, finding that accreting WDs can account for the unresolved soft X‑ray emission observed in nearby galaxies and contribute significantly to the integrated background at keV energies \cite{Chen2015}. These models incorporate binary evolution codes and grids of WD evolutionary tracks, allowing estimates of the combined X‑ray output and its dependence on accretion physics and star‐formation history.

Recent high‑resolution measurements of the Galactic hard X‑ray and soft $\gamma$-ray background, such as those based on long‑term INTEGRAL/IBIS observations, provide new constraints on the spectral shape of this emission and reinforce the interpretation that unresolved accreting WDs dominate the X‑ray background below $\sim60$~keV \cite{Krivonos:2006px, Krivonos:2024bih}. 
This component typically exhibits a spectral minimum around $\sim80$~keV, above which truly diffuse $\gamma$‑ray processes (from CR electrons) contribute more substantially. By combining these observational insights with physically motivated population models, it becomes possible to construct a comprehensive model of the accreting WD contribution to the Galactic X‑ray background that can be tested against broad‑band X‑ray observations and used to separate this stellar background from other diffuse emission processes.

\section{Analysis}

In this section, we explain first how the theoretical X-ray fluxes coming from PBH evaporation, CR $e^\pm$ and accreting WDs are evaluated in order to be compared with the INTEGRAL/IBIS observations. Second, we describe how these theoretical fluxes are fitted to the observations, in order to derive the desired upper limits on $f_{\rm PBH}$.

\subsection{Extraction of the modeled X-ray emissions}

In order to perform a robust comparison between the predicted X-ray emission, from both DM and astrophysical sources, and the data, we adopt the following procedure. For each ScW used to construct the longitudinal profiles and energy spectra, we evaluate the predicted X-ray flux, convolve it with the IBIS/ISGRI collimator response function, and average the resulting predictions over all relevant ScWs. We then convert our predictions from physical flux units to Crab units using the fact that the Crab Nebula emission spectrum can be modeled as a power law $d\Phi_\gamma^{\rm Crab}/dE_\gamma = 10\,(E_\gamma/\textrm{keV})^{-2.1}$ cm$^{-2}$ s$^{-1}$ keV$^{-1}$~\cite{Churazov:2006bk}.

The characterization of the collimator response function described in Sec.~\ref{sec:data} is an approximation. In principle, the response function is not axisymmetric, and the satellite roll angle should be taken into account when calibrating the response using observations of the Crab Nebula. We refer the reader to Fig.~4 of Ref.~\cite{Krivonos:2006px} for an empirical estimate of the collimator response in two energy bands. However, the simpler procedure described in Sec.~\ref{sec:data} was used in Ref.~\cite{Krivonos:2024bih} to compute the longitudinal profiles and energy spectra. We therefore adopt the same procedure in our baseline analysis, ensuring a consistent comparison between the predicted X-ray emission and the data.

%Since the characterization of the collimator response function can be a source of systematic uncertainty, we also use an alternative procedure and estimate its impact on our results. Within the fully coded region, instead of performing a regression between the angular separation, defined as the angle between the Crab Nebula and the center of the ScW, and the Crab detector count rate independently of the revolution number, as done in Ref.~\cite{Krivonos:2024bih}, we perform a regression between the angular separation and the ratio of the Crab detector count rate to the modeled Crab count rate at the corresponding revolution. This second procedure removes the leading time dependence associated with the degradation of ISGRI over the mission, and therefore provides a useful cross-check of the baseline collimator-response model.
Fig.~\ref{fig:PBHmap} shows the total X-ray intensity expected from the evaporation of $10^{16}$~g PBHs before (top panel) and after (bottom panel) convolution with the IBIS detector response functions. A clear imprint of the ISM gas can be seen in the spatial distribution of electrons injected by the PBHs, which is suppressed along the Galactic plane because of the strong energy losses arising from interactions with the gas, primarily through ionization of neutral hydrogen.
Contour lines indicating the enclosed fraction of FSR emission are overlaid on both maps. In the non-convolved map (top panel), these contours closely trace the underlying DM distribution. After convolution with the instrument response functions, however, the FSR emission appears significantly more extended, clearly illustrating the spatial smearing introduced by the detector's angular resolution.

\begin{figure*}[t]
    \centering
    \includegraphics[width=\linewidth]{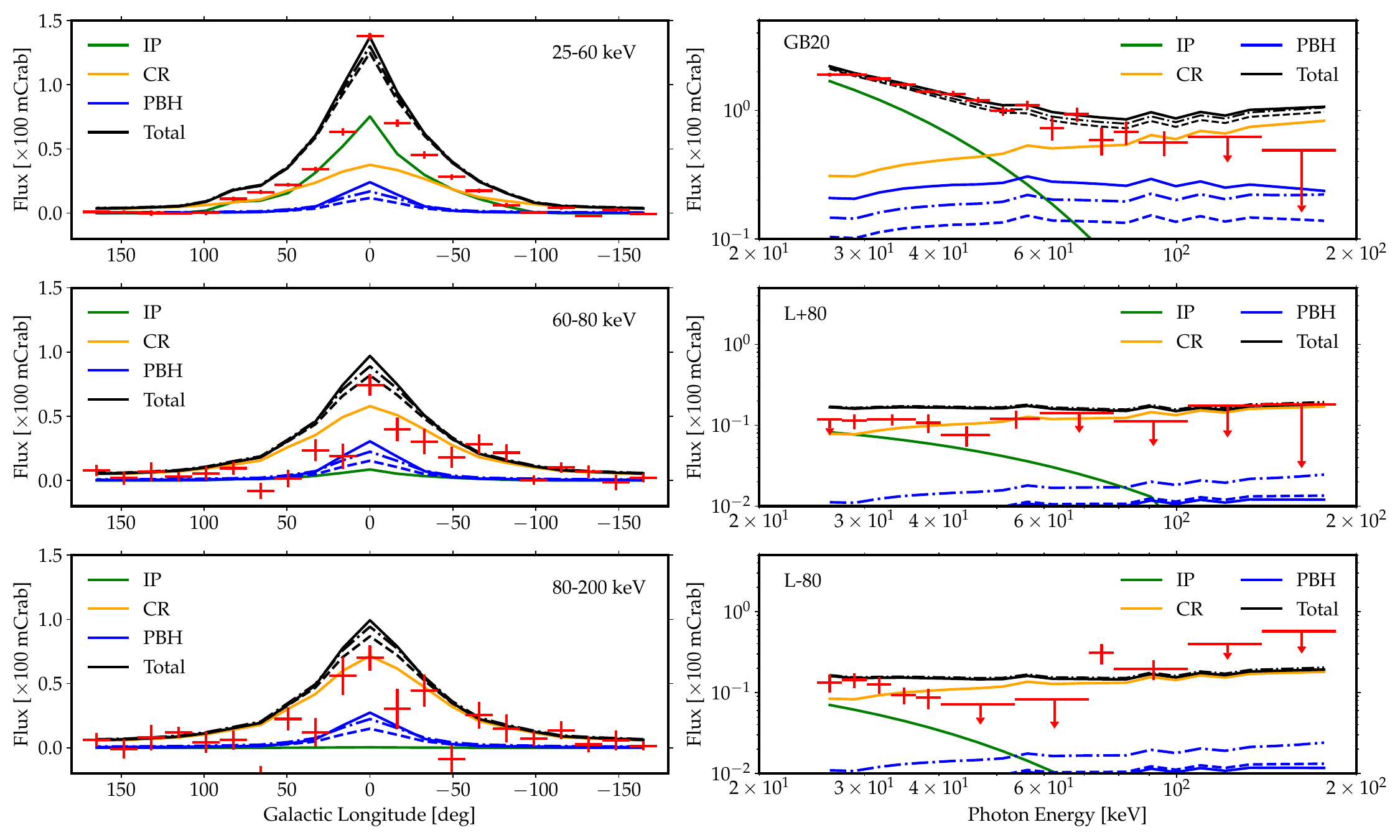}
    \caption{Morphology and spectra of the X-ray emissions, the red points being the measurements from IBIS/ISGRI. The PBH-induced emissions (in blue) are from three cases: $M_{\rm PBH}=10^{15}$g and $f_{\rm PBH}=3\times10^{-7}$ (solid lines), $M_{\rm PBH}=10^{16}$g and $f_{\rm PBH}=3\times10^{-4}$ (dashed), and $M_{\rm PBH}=10^{17}$g and $f_{\rm PBH}=0.3$ (dot-dashed). The predicted CR contribution is in orange, the IP contribution from WDs in green and the sum of all components is shown in black.}
    \label{fig:fluxvsmPBH}
\end{figure*}

Fig.~\ref{fig:fluxvsmPBH} compares the morphology and spectra of the extracted X-ray observations with the expected background components discussed above. The WD contribution (labeled IP in the legend) is shown in green, while the CR component is shown in orange. The spectral panels clearly illustrate that the WD emission dominates at low energies, whereas the predicted CR contribution provides a good description of the observed emission above a few tens of keV.
For comparison, we also show the morphology and spectra expected from the IC emission associated with PBH evaporation in blue, where the different line styles correspond to different PBH masses (and different associated values of $f_{\rm PBH}$). The predicted PBH morphologies and spectra reflect the interplay between the energy of the injected electrons, their diffusion, and their energy losses in the ISM.

It is also worth noting that the CR model already provides a reasonably good description of the high-energy observations without any tuning. Nevertheless, as discussed below, our fitting procedure allows the normalizations of both the WD and CR components to vary freely. This approach enables us to determine which background component best reproduces the observed emission, assess whether an additional X-ray contribution from PBH evaporation is statistically preferred, and derive robust upper limits on the PBH fraction.

\subsection{Fitting procedure}

To summarize, we fit the total X-ray emission with three modeled contributions. The first is the emission induced by CR $e^\pm$, described by a single normalization parameter $A_{\rm CR}$, common to all three sky regions and all energy bands, as discussed in Sec.~\ref{subsec:crel}.

The second contribution is the emission from accreting WDs, discussed in Sec.~\ref{subsec:WDs}. Following Ref.~\cite{Krivonos:2024bih}, we model the spectrum up to 100 keV of this component using the X-ray spectrum of intermediate polars (IPs), which are expected to be the dominant emitters among accreting WDs, from Ref.~\cite{Suleimanov:2004aq}. This model depends on the average WD mass, $M_{\rm WD}$ (with the allowed range $0.3-1.4\,M_\odot$), and on an overall normalization, $A_{\rm IP}$. We allow the WD mass to vary independently in the three sky regions, while the IP normalization is treated as described below.

For the longitudinal morphology, we use the fact that the GRXE intensity is known to follow the Galactic near-infrared (NIR) brightness~\cite{Krivonos:2006px,2006A&A...452..169R}, which traces the stellar mass distribution. We therefore use the NIR intensity profile measured by COBE/DIRBE at 4.9 $\mu$m, convolved with the IBIS/ISGRI collimator response, as shown in Fig.~1 of Ref.~\cite{Krivonos:2024bih}. We allow this morphological template to have an independent normalization $A_{\rm NIR}^{\Delta E}$ in each of the three broad energy bands.

The NIR and IP descriptions are not treated as independent components. Instead, they are tied together by requiring that, for a given energy band and sky region, the region-averaged NIR contribution to the longitudinal profile reproduces the corresponding band averaged IP flux in the energy spectrum. Schematically, for a sky region and an energy band, this condition reads
\begin{equation}
    \frac{A_{\rm IP}^\textrm{reg}}{\Delta E} \int_{\Delta E} dE\, \Phi_{\rm IP}(E;M_{\rm WD}^\textrm{reg}) = \frac{A_{\rm NIR}^{\Delta E}}{\Delta \ell_\textrm{reg}} \int_{\Delta \ell_\textrm{reg}} d\ell\, \Phi_{\rm NIR}^{\Delta E}(\ell)\;.
\end{equation}
In practice, for a fixed set of WD masses, one for each sky region, we first determine the three NIR normalizations $A_{\rm NIR}^{\Delta E}$ from the \texttt{GB20} region. These normalizations are fixed by matching the NIR morphology to the IP spectrum in \texttt{GB20}, whose amplitude is controlled by $A_{\rm IP}^{\texttt{GB20}}$ and whose shape depends on $M_{\rm WD}^{\texttt{GB20}}$. We then use the same matching condition in the 25$–$60 keV band to determine the IP normalizations in the two outer regions, $A_{\rm IP}^{\texttt{L-80}}$ and $A_{\rm IP}^{\texttt{L+80}}$. Thus, the accreting-WD component is fully specified by $A_{\rm IP}^{\texttt{GB20}}$ and the three WD masses, $M_{\rm WD}^{\texttt{GB20}}$, $M_{\rm WD}^{\texttt{L-80}}$, and $M_{\rm WD}^{\texttt{L+80}}$.

The third contribution is the PBH evaporation signal. For the analysis considered here, this contribution is the sum of the prompt FSR emission and the ICS emission induced by the injected $e^\pm$, as discussed in Sec.~\ref{sec:evap}. Its normalization is controlled by the PBH abundance $f_{\rm PBH}$, while its shape depends on the PBH mass $M_{\rm PBH}$. The total model is therefore fitted with the parameter set $\boldsymbol{\theta}=(f_{\rm PBH},A_{\rm CR},A_{\rm IP}^{\texttt{GB20}},M_{\rm WD}^{\texttt{GB20}},M_{\rm WD}^{\texttt{L-80}},M_{\rm WD}^{\texttt{L+80}})$, by scanning over a range of $M_{\rm PBH}$, and where $f_{\rm PBH},A_{\rm CR}$, and $A_{\rm IP}^{\texttt{GB20}}$ are sampled logarithmically, while the WD masses $M_{\rm WD}^{\rm reg}$ are sampled linearly. The derived quantities $A_{\rm NIR}^{\Delta E}$, $A_{\rm IP}^{\texttt{L-80}}$, and $A_{\rm IP}^{\texttt{L+80}}$ are recomputed for each point in this parameter space.

Since our primary objective is to derive upper limits on $f_{\rm PBH}$, we treat $A_{\rm CR}$, $A_{\rm IP}^{\texttt{GB20}}$, $M_{\rm WD}^{\texttt{GB20}}$, $M_{\rm WD}^{\texttt{L-80}}$, and $M_{\rm WD}^{\texttt{L+80}}$ as nuisance parameters, %which are marginalized over in the Bayesian analysis and profiled over in the frequentist analysis. 
These parameters can be partially degenerate with the PBH contribution: variations in the normalizations or spectral shapes of the astrophysical components can either mimic or absorb part of a putative PBH signal.

Nevertheless, external information motivates physically informed constraints on some of these parameters. The nominal CR template corresponds to $A_{\rm CR}=1$, although uncertainties in the CR electron distribution, propagation model, and predicted emission motivate allowing its normalization to vary. Similarly, previous studies of the Galactic ridge emission find characteristic WD masses of approximately $0.5-0.66\,M_\odot$~\cite{Krivonos:2006px,Turler:2010pm,Yuasa:2012qe,Heard:2012pt,Perez:2019bgo}.

To assess the sensitivity of the inferred PBH limits to the treatment of these nuisance parameters, we consider three scenarios:

\textbf{Free priors}: We impose no informative constraints on the nuisance parameters beyond broad physical or numerical bounds. Specifically, we adopt uniform priors in the variables used in the fit. Since $A_{\rm CR}$ and $A_{\rm IP}^{\texttt{GB20}}$ are sampled logarithmically, these choices correspond to log-uniform priors on their physical normalizations. These priors therefore do not preferentially select their nominal values within the allowed ranges.

$\mathbf{A_{\rm{\textbf{CR}}}\,\&\,M_{\rm{\textbf{WD}}}}$ \textbf{penalized}: We constrain $A_{\rm CR}$ to remain close to unity and the three effective WD masses to remain close to $0.6\,M_\odot$. In the implementation, the corresponding log-prior term is
    \begin{equation}
        -2\log \mathcal{\pi}_{\rm pen}(\boldsymbol{\theta})=\left(\frac{\log A_{\rm CR}}{\log(1+\sigma_{\rm pen})}\right)^2+\sum_{\rm reg}\left(\frac{M_{\rm WD}^{\rm reg}-0.6M_\odot}{\sigma_{\rm pen} M_\odot}\right)^2
    \end{equation}
    where we chose $\sigma_{\rm pen}=0.15$. The normalization $A_{\rm IP}^{\texttt{GB20}}$ remains subject only to its free prior.

$\mathbf{A_{\rm{\textbf{CR}}}=1,M_{\rm{\textbf{WD}}}}$ \textbf{penalized}: As a limiting case, we fix $A_{\rm CR}=1$, thereby removing the CR normalization as a free parameter, while retaining the Gaussian constraints on the three WD masses. This scenario tests the impact of assuming that the CR template has the correct absolute normalization.

To assess the robustness of our results with respect to the statistical construction, we perform both Bayesian and frequentist analyses. If $\boldsymbol{\Phi}_{\rm data}$ and $\boldsymbol{\Phi}_{\rm model}(\boldsymbol{\theta})=\boldsymbol{\Phi}_{\rm PBH}(f_\textrm{PBH})+\boldsymbol{\Phi}_{\rm IP}(A_{\rm IP},M_{\rm WD}^{\rm regs})+\boldsymbol{\Phi}_{\rm CR}(A_{\rm CR})$ are respectively the data and model prediction vectors (containing all of the 126 longitudinal/spectral bins). In this case, the log-likelihood is defined as (up to an additive constant independent of the model parameters)
\begin{equation}
    -2\log \mathcal{L}(\boldsymbol{\theta})=\mathbf{R}(\boldsymbol{\theta})^\mathrm{T} \mathbf{C}^{-1}\mathbf{R}(\boldsymbol{\theta})\;,
\end{equation}
where $\mathbf{R}(\boldsymbol{\theta})=\boldsymbol{\Phi}_{\rm data}-\boldsymbol{\Phi}_{\rm model}(\boldsymbol{\theta})$ is the vector of residuals and $\mathbf{C}$ is the covariance matrix, as defined in Sec.~\ref{sec:data}. When correlations between the measurements are neglected, then $\mathbf{C}=\mathrm{diag}(\sigma_i^2)$, and the log-likelihood reduces to
\begin{multline}
    -2\log \mathcal{L}(\boldsymbol{\theta})=\sum_{\rm{reg}}\sum_i^{E\textrm{-bins}}\left(\frac{\Phi_{{\rm model},i}^{\rm reg}(\boldsymbol{\theta})-\Phi_{{\rm data},i}^{\rm reg}}{\sigma_{{\rm data},i}^{\rm reg}}\right)^2+\\
    + \sum_{\rm{band}}\sum_i^{\ell\textrm{-bins}}\left(\frac{\Phi_{{\rm model},i}^{\rm band}(\boldsymbol{\theta})-\Phi_{{\rm data},i}^{\rm band}}{\sigma_{{\rm data},i}^{\rm band}}\right)^2
\end{multline}
The treatment of the nuisance parameters depends on the chosen aforementioned prior scenario. In addition, we impose hard bounds on the parameter space explored by the MCMC: $-15<\log f_{\rm PBH}<0$, as PBH cannot constitute more than 100\% of dark matter, $-2<\log A_{\rm CR}<2$, i.e.~the normalization of the CR background can vary up to two order of magnitude from the benchmark template, $0.3<M_{\rm WD}<1.4$, which is the range allowed by the IP model, and $-10<\log A_{\rm IP}^{\texttt{GB20}}<2$. In the Bayesian analysis, the log-posterior distribution is therefore
\begin{equation}
\log p\left(\boldsymbol{\theta}|\boldsymbol{\Phi}_{\mathrm{data}}\right)=\log\mathcal{L}(\boldsymbol{\theta})+\log\pi_\textrm{pen}(\boldsymbol{\theta})\;.
\end{equation}

In the frequentist analysis, the corresponding Gaussian constraints are incorporated as additive penalty terms. We consequently define the effective chi-square as $\chi_{\mathrm{eff}}^2(\boldsymbol{\theta})=-2\log p\left(\boldsymbol{\theta}|\boldsymbol{\Phi}_{\mathrm{data}}\right)$

We sample the posterior distribution using the Markov chain Monte Carlo (MCMC) package \texttt{emcee}~\cite{Foreman-Mackey:2012any}. An exploratory run, with walkers initialized throughout the allowed prior volume, is first used to identify separated regions of high posterior density. The production walkers are then initialized around several of these regions. The production run is continued until its length exceeds 100 integrated autocorrelation times for every parameter and the estimated autocorrelation times vary by less than $1\%$. The burn-in samples are discarded, and the remaining chains are thinned according to their autocorrelation times.

We report the sampled maximum-a-posteriori point as the Bayesian best-fit parameter set. The one-sided $95\%$ Bayesian upper limit on $f_{\rm PBH}$ is defined as the 95th percentile of its marginalized posterior distribution.

For the frequentist analysis, we first minimize $\chi_{\mathrm{eff}}^2$ over the complete parameter space using the differential-evolution algorithm~\cite{Storn:1997uea}, as implemented in \texttt{scipy}~\cite{2020SciPy-NMeth}. This procedure determines the global minimum
\begin{equation}\chi_{\min}^2=\min_{f_{\mathrm{PBH}},\boldsymbol{\eta}}\chi_{\mathrm{eff}}^2\left(f_{\mathrm{PBH}},\boldsymbol{\eta}\right)\;,
\end{equation}
where $\boldsymbol{\eta}$ denotes the nuisance parameters. We then construct the profile chi-square by fixing $f_{\mathrm{PBH}}$ and minimizing over the nuisance parameters:\begin{equation}
    \chi_{\mathrm{prof}}^2(f_{\mathrm{PBH}})=\min_{\boldsymbol{\eta}}\chi_{\mathrm{eff}}^2\left(f_{\mathrm{PBH}},\boldsymbol{\eta}\right)\;.
\end{equation}
The nuisance-parameter minimization is repeated from several initial points to reduce the risk of convergence to a local minimum. We define the profile-likelihood test statistic as
\begin{equation}
    \Delta\chi^2(f_{\mathrm{PBH}})=\chi_{\mathrm{prof}}^2(f_{\mathrm{PBH}})-\chi_{\min}
\end{equation}
Under the standard asymptotic approximation for one parameter of interest ($f_{\mathrm{PBH}}$), the one-sided $95\%$ upper confidence limit is given by the upper solution of $\Delta\chi^2\left(f_{\mathrm{PBH}}\right)=2.71$.

Several sources of systematic uncertainties are expected to impact our results. They can be model related: choice of the DM profile, PBH mass distribution, black-hole geometry, and propagation model parameters (e.g.~the Alfvén speed $v_A$). The other sources of systematic uncertainties are related to the fitting procedure. In our analysis, these are: the choice of the prior on the nuisance parameter, the estimation of the covariance matrix, and the chosen statistical inference.

We adopt the following benchmark: NFW profile, monochromatic and Schwarzschild PBHs, $v_A=13.4$ km/s, $A_{\rm CR}\,\&\,M_{\rm WD}$ penalized prior, frequentist inference, no covariance. This choice is a representative median case of all of the considered sources of systematic uncertainties.

\section{Results}

As a first test, we perform the Bayesian analysis for the background-only hypothesis. The results are summarized in Tab.~\ref{tab:noPBH}. As reported in the table, adding the covariance matrix in the analysis allow the best-fit $\chi_{\rm eff}^2$/d.o.f.~to decrease by a factor $\sim 2$, and finds similar best-fit parameters.
We remind the reader that the WD masses are fitted independently in each of the regions considered, allowing for possible spatial variations in the WD population. Although the average WD mass is expected to vary smoothly across the Galaxy, differences between the Galactic Bulge and the outer Galactic regions are anticipated due to their distinct stellar populations and formation histories. Furthermore, asymmetries in the signal could naturally arise from spatial variations in the WD population rather than from the fitting procedure itself.

\begin{table}[t]
    \centering
    \begin{tabular}{|c|c|c|c|}
        \hline
         Prior case & $\chi^2_{\rm eff}$/d.o.f. & $A_{\rm CR}$ & $A_{\rm IP}\times10^{-6}$ \\
         \hline
         Free priors & $4.86|2.48$ & $0.19|0.14$ & $3.13|3.12$ \\
         $A_{\rm CR}\,\&\,M_{\rm WD}$ pen. & $5.48|3.02$ & $0.39|0.44$ & $4.30|3.82$ \\
         $A_{\rm CR}=1,M_{\rm WD}$ pen. & $8.49|4.47$ & $1.00|1.00$ & $11.4|6.45$ \\
         \hline
    \end{tabular}
    \par\medskip
    \begin{tabular}{|c|c|c|c|}
        \hline
         Prior case & $M_{\rm WD}^\texttt{GB20}$ ($M_\odot$) & $M_{\rm WD}^\texttt{L+80}$ ($M_\odot$) & $M_{\rm WD}^\texttt{L-80}$ ($M_\odot$)\\
         \hline
        Free priors & $0.71|0.71$ & $0.99|0.98$ & $0.46|0.45$ \\
        $A_{\rm CR}\,\&\,M_{\rm WD}$ pen. & $0.65|0.66$ & $0.82|0.77$ & $0.48|0.47$ \\
        $A_{\rm CR}=1,M_{\rm WD}$ pen. & $0.50|0.56$ & $0.78|0.73$ & $0.55|0.49$ \\
         \hline
    \end{tabular}
    \caption{Best-fit $\chi_{\rm eff}^2$/d.o.f. and parameters in the background only hypothesis, and for the three explored prior scenarios. In each column, the values obtain with (without) the estimated covariance matrix are shown on the right (left). }
    \label{tab:noPBH}
\end{table}

\begin{figure}[t]
    \centering
    \includegraphics[width=\linewidth]{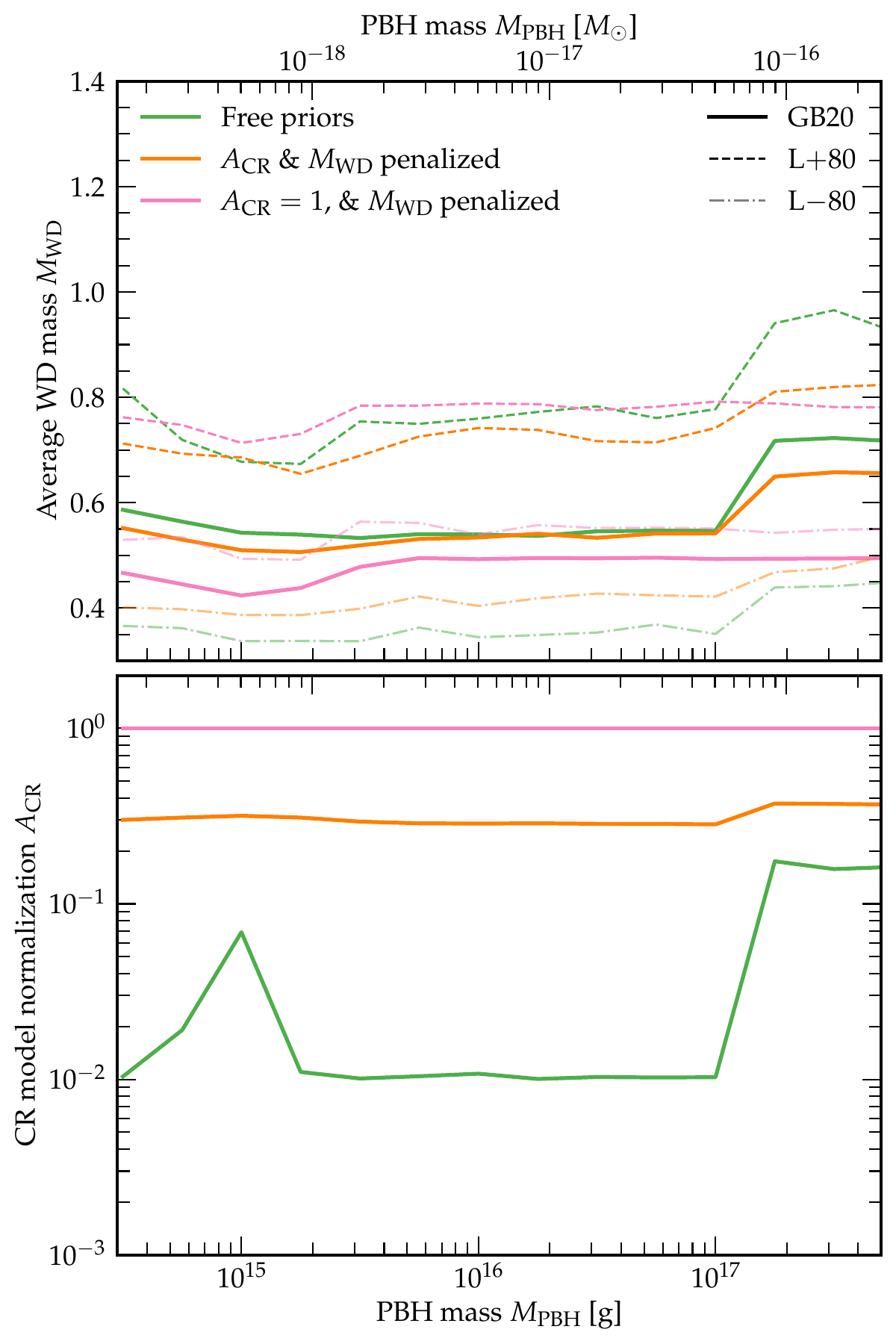}
    \caption{Best-fit values of the nuisance parameters as a function of the PBH mass for the three prior scenarios considered in the analysis. Top panel: effective WD masses in the three sky regions, $M_{\rm WD}^{\texttt{GB20}}$, $M_{\rm WD}^{\texttt{L-80}}$, and $M_{\rm WD}^{\texttt{L+80}}$ studied here. Bottom panel: best-fit normalization of the CR contribution, $A_{\rm CR}$. The solid lines indicate the best-fit values obtained from the global analysis.}
    \label{fig:fit_MWD_ACR}
\end{figure}

We show in Fig.~\ref{fig:fit_MWD_ACR} the best-fit values obtained for the WD mass, $M_{\rm WD}$, and the normalization of the CR component, $A_{\rm CR}$, as a function of the PBH mass, for the three prior choices and regions considered in our analysis. As shown, the best-fit values of $M_{\rm WD}$ remain in the range $0.4$--$0.8\,M_\odot$. We note that although WD masses below $\sim0.5\,M_\odot$ cannot be produced through standard single-star evolution within the age of the Universe, they might be the outcome of binary evolution~\cite{Marsh1995, Zhang2018}. %Although a small population of apparently single low-mass WDs has been identified, their origin is likewise believed to involve non-standard evolutionary channels, such as binary interactions or mergers~\cite{Zhang2018}.
Meanwhile, for the case of A$_{\rm CR}$, the fits tend to drive this parameter below unity. However, in the free-prior case, we find a preferred reduction of nearly two orders of magnitude. Such a strong suppression appears difficult to justify, given that the model provides a satisfactory description of the fluxes in the MeV band. Although variations of a factor of a few may be expected due to our limited understanding of low-energy CR electrons, a suppression of this magnitude seems unlikely. This suggests that the fit favors a scenario in which the high-energy part of the X-ray diffuse flux investigated here is predominantly explained by the PBH contribution, with a subdominant CR electron component.

\begin{figure*}[t]
    \centering
    \includegraphics[width=\linewidth]{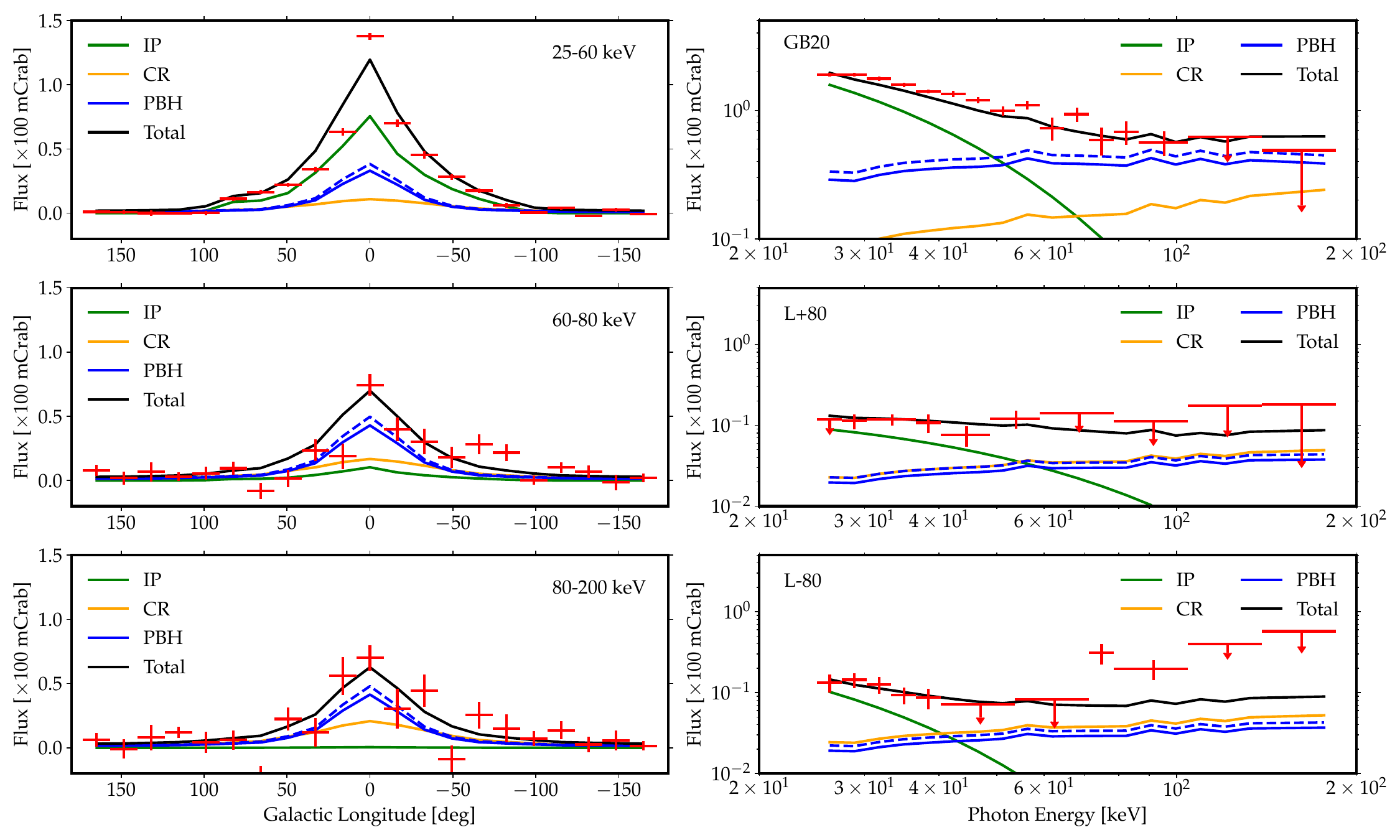}
    \caption{Best-fit morphology and spectra for $M_{\rm PBH}=10^{16}$ g. The solid lines show the total emission corresponding to the best-fit PBH fraction, $f_{\rm PBH}=8.37\times10^{-4}$, while the dashed lines indicate the $95\%$ C.L. upper limit, $f_{\rm PBH}^{95\%}=9.69\times10^{-4}$. The corresponding best-fit astrophysical parameters are $A_{\rm CR}=0.29$, $A_{\rm IP}=8.11\times10^{-6}$, and $(M_{\rm WD}^{\texttt{GB20}},M_{\rm WD}^{\texttt{L+80}},M_{\rm WD}^{\texttt{L-80}})=(0.53,0.72,0.41)\,M_{\odot}$, with a goodness of fit of $\chi^2/{\rm d.o.f.}=4.74$. The individual contributions from PBH evaporation, CR electrons, and accreting WDs (IP) are shown with the colors indicated in the legend.}
    %\caption{$M_{\rm PBH}=10^{16}$g and best-fit $f_{\rm PBH}=8.37\times10^{-4}$ (solid), its $95\%$ upper limit $f_{\rm PBH}^{95\%}=9.69\times10^{-4}$ (dashed), and $A_{\rm CR}=0.29$, $A_{\rm IP}=8.11\times10^{-6}$, and $(M_{\rm WD}^{\texttt{GB20}},M_{\rm WD}^{\texttt{L+80}},M_{\rm WD}^{\texttt{L-80}})=(0.53,0.72,0.41)\,M_{\odot}$, for $\chi^2/\rm{d.o.f.}=4.74$.}
    \label{fig:fluxBF}
\end{figure*}

Likewise, Fig.~\ref{fig:fluxBF} shows the best-fit morphology and spectra for a PBH mass of $M_{\rm PBH}=10^{16}$~g. The figure also includes a dashed line indicating the corresponding $95\%$ C.L. upper limit on the PBH contribution. Interestingly, the best-fit PBH signal becomes the dominant component at high energies in all regions and is most prominent toward the Galactic Center. As a result, the normalization of the CR component is substantially reduced, highlighting a partial degeneracy between these two contributions. However, the fit favors a lower CR normalization, primarily driven by the inner Galaxy observations, where the inclusion of a PBH component provides a better description of the observed emission.

\begin{figure}[t]
    \centering
    \includegraphics[width=\linewidth]{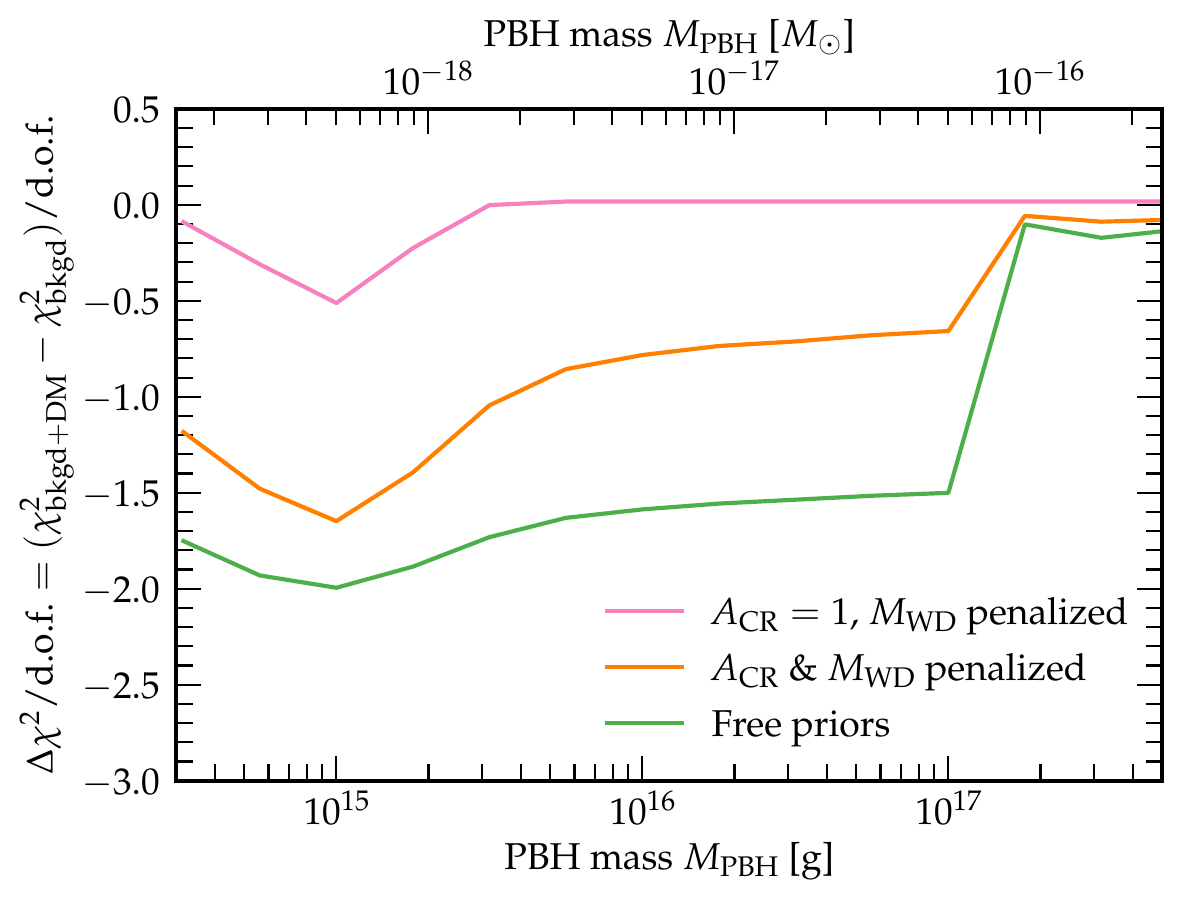}
    \caption{Difference in the goodness of fit between the PBH+background and background-only hypotheses, per degree of freedom. The curves correspond to the three prior scenarios considered in the analysis, including correlations between observations. Negative values indicate that the PBH hypothesis provides a better fit.}
    %\caption{$\chi^2_{\rm bkgd}/{\rm d.o.f.}=2.48 (4.86)$ for the free prior, $\chi^2_{\rm bkgd}/{\rm d.o.f.}=3.02 (5.48)$ for the $A_{\rm CR}\,\&\,M_{\rm WD}$ penalized, $\chi^2_{\rm bkgd}/{\rm d.o.f.}=4.47 (8.48)$ for the $A_{\rm CR}=1,\,M_{\rm WD}$ penalized with (without) covariance}
    \label{fig:chi2}
\end{figure}

Remarkably, for all combinations of fits and priors considered, we find a significant reduction in the $\chi^2$ value when including the PBH contribution compared to the background-only case (see Fig.~\ref{fig:chi2}). This indicates that the astrophysical components considered may not fully account for the observed emission, and that the data favor the presence of an additional component (potentially with a morphology concentrated towards the Galactic Center). Such a contribution could originate from unmodeled astrophysical sources, or from an exotic component associated with PBHs or DM. However, identifying the nature of this additional component from the current data alone remains challenging, given the degeneracies between its spectral shape and spatial distribution and those of the diffuse Galactic emission. The preference for an additional component therefore highlights that our understanding of the hard X-ray background at these energies may still be incomplete, and improved modeling of Galactic diffuse emission will be essential for both identifying potential new physics signals and deriving robust constraints on exotic contributions.
In this way, we emphasize that this improvement in the fit should not be interpreted as evidence for a PBH signal. %Given the remaining uncertainties in the modeling of the diffuse Galactic X-ray emission, the additional component favored by the fit could also arise from unaccounted astrophysical processes. 
Consequently, we adopt a conservative approach and use the data to derive upper limits on the allowed PBH contribution rather than claim a detection.

\begin{figure}[t]
    \centering
    \includegraphics[width=\linewidth]{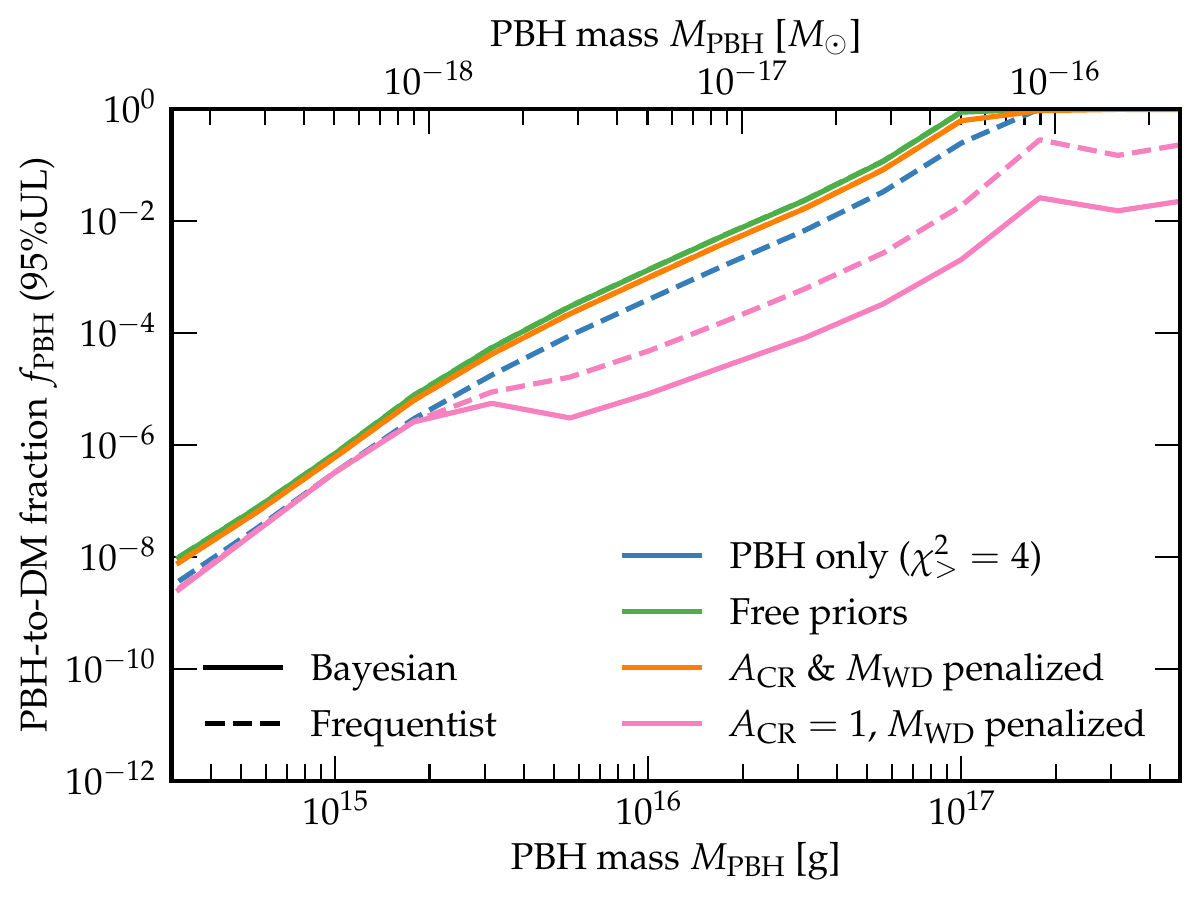}
    \caption{Impact of the choice of nuisance-parameter priors on the derived PBH constraints. Solid lines are obtained when adopting our Bayesian, MCMC, analysis and dashed lines when considering the frequestist approach described above. The three colors shown correspond to free priors (green), $A_{\rm CR}\,\&\,M_{\rm WD}$ penalized priors (orange), and the limiting case of fixed $A_{\rm CR}=1$ with penalized $M_{\rm WD}$ (pink). Our benchmark analysis corresponds to the $A_{\rm CR}\,\&\,M_{\rm WD}$ penalized scenario and a frequentist approach. The DM only case was derived with a conservative method, considering that all the data points are upper limits (see e.g.~Ref.~\cite{Cirelli:2023tnx} for the full statistical method).}
    \label{fig:PBH_prior}
\end{figure}

Our constraints on the PBH fraction are shown in Fig.~\ref{fig:PBH_prior} for all the priors considered and for both the Bayesian and frequentist approaches tested. Although the two methods yield very similar constraints, small differences are present. %, likely arising from the difference between the treatment of the background components. 
In the Bayesian approach, the constraints are derived from a Monte Carlo sampling of the posterior probability distribution, which is not necessarily Gaussian. In contrast, in the frequentist approach we derive upper limits by determining the value of the PBH fraction for which the $\chi^2$ increases by 2.71 with respect to its minimum, corresponding to the standard one-sided 95\% confidence level. We note that the DM-only case (derived using the same conservative test as in e.g. Ref.~\cite{Cirelli:2023tnx}) almost overlaps with the case in which the background parameters are left free (Free priors). Therefore, the spread between these curves can be interpreted as an estimate of the uncertainty associated with the treatment of the astrophysical backgrounds. The small size of this uncertainty indicates that our constraints are relatively robust against reasonable variations in the background modeling.
 
\begin{figure}[t]
    \centering
    \includegraphics[width=\linewidth]{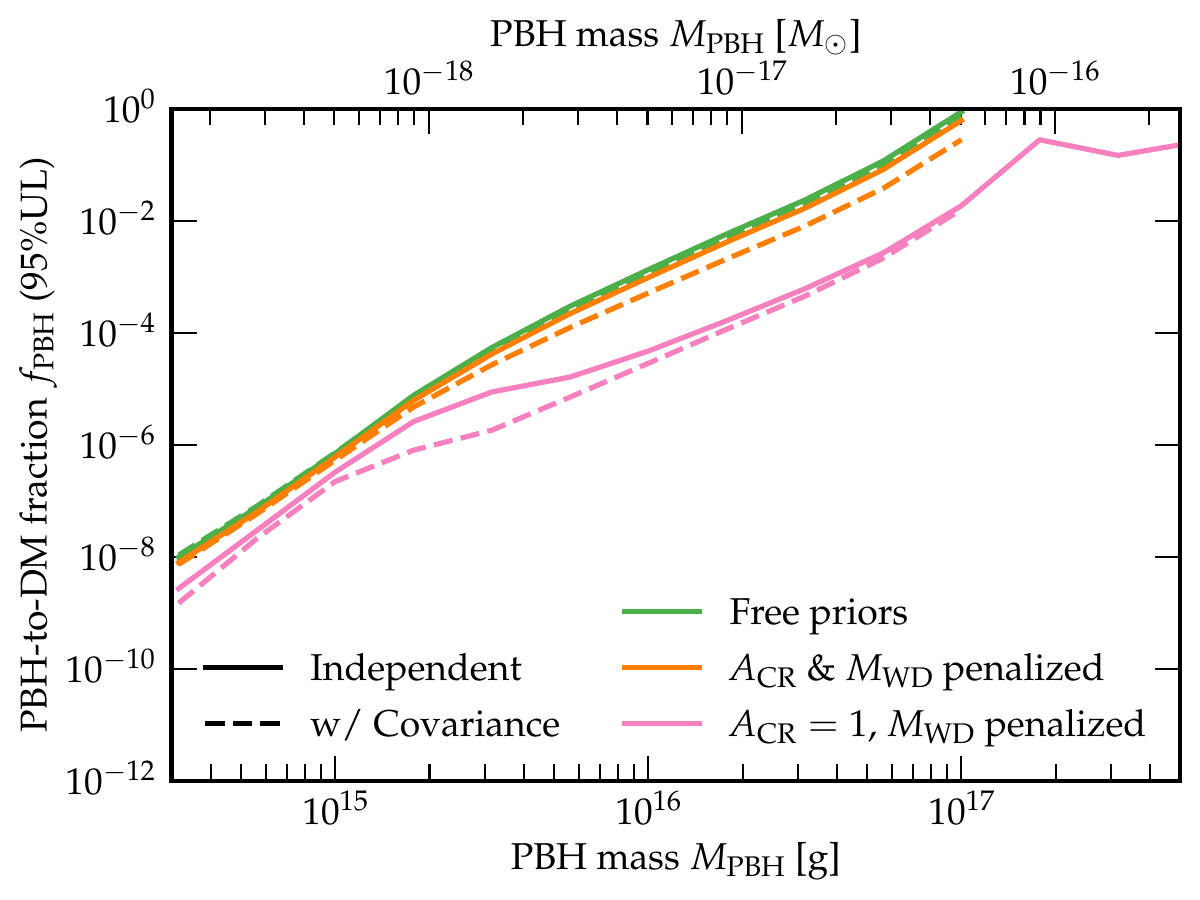}
    \caption{Impact of including correlations between observations on the derived PBH constraints. The constraints obtained with and without covariance are shown as dashed and solid lines, respectively, using the frequentist approach adopted in the analysis.}
    \label{fig:PBH_cov}
\end{figure}

Similarly, Fig.~\ref{fig:PBH_cov} illustrates the impact of accounting for the correlations between observations within the same observing revolution in the frequentist analysis, which generally yields slightly more conservative limits. As can be seen, these correlations have a relatively small effect on the resulting constraints. Nevertheless, accounting for them is important, and this analysis therefore provides our most complete and robust benchmark.

\begin{figure}[t]
    \centering
    \includegraphics[width=\linewidth]{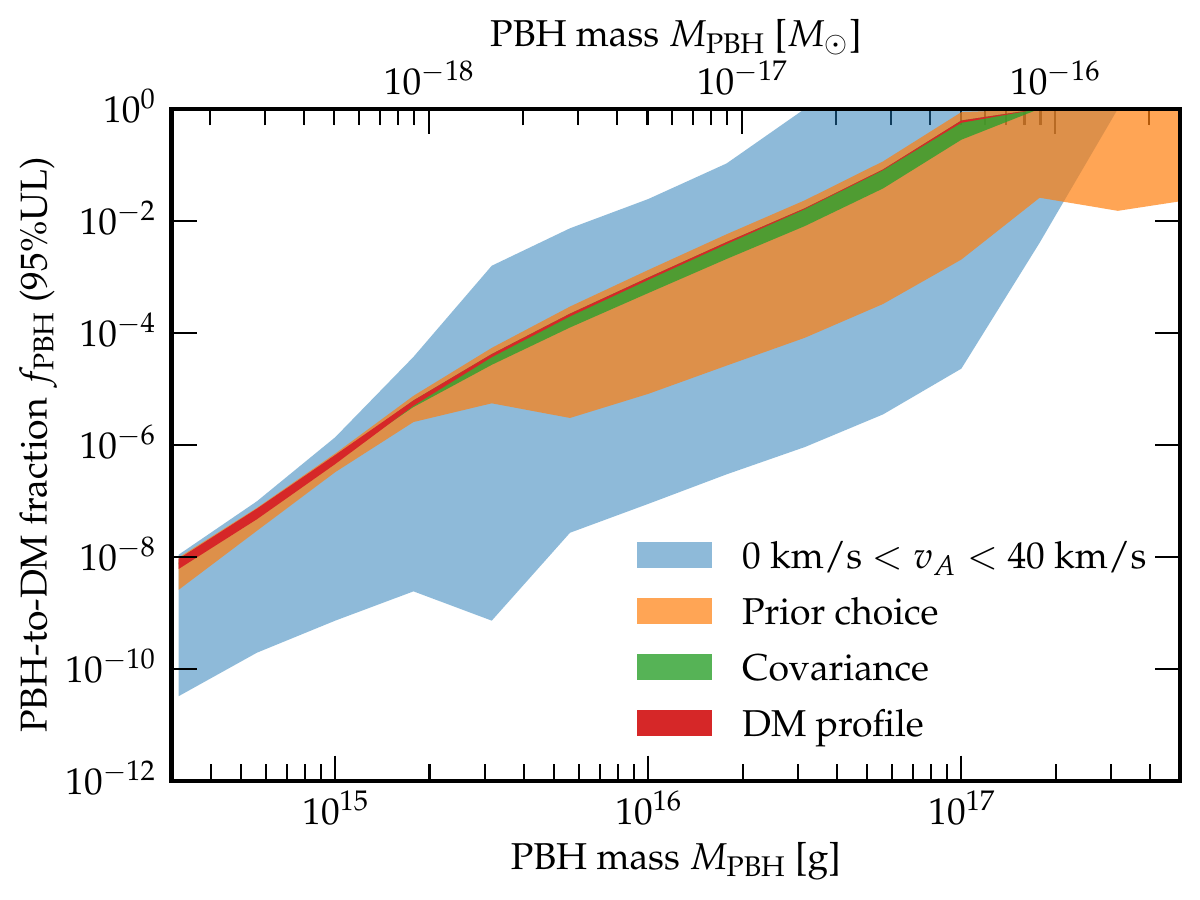}
    \caption{Impact of astrophysical and analysis uncertainties on the derived PBH constraints. The bands show the variations induced by the main sources of uncertainty considered in the analysis: CR propagation (different reacceleration scenarios parametrized by $v_A$), the Galactic DM density profile (from $\gamma=0$, i.e.~isothermal, to $\gamma=1.5$), covariance between observations, and the choice of nuisance-parameter priors.}
    \label{fig:PBH_unc}
\end{figure}

Moreover, we assess the impact of additional sources of uncertainty on our constraints. In particular, Fig.~\ref{fig:PBH_unc} illustrates the uncertainty associated with the choice of DM density profile. We find that the resulting variation in the constraints remains within a few tens of percent. This relatively small impact can be understood for several reasons. First, the PBH-induced signal scales only linearly with the DM density. Second, the electrons responsible for the X-ray emission propagate over angular scales of several degrees before losing their energy, effectively smoothing differences between DM profiles. Finally, the constraints are driven primarily by data at relatively large Galactic longitudes, where the WD and CR contributions are less dominant than in the GB and where variations among the different DM profiles have a smaller effect on the predicted signal. 
However, the dominant astrophysical uncertainty affecting our constraints arises from the propagation of the electrons injected by PBHs. In particular, the largest source of uncertainty is the effect of diffusive reacceleration, as previously discussed in Refs.~\cite{DelaTorreLuque:2024qms, DelaTorreLuque:2023olp, Balaji:2025afr, DelaTorreLuque:2023nhh}. This effect is parametrized through the Alfvén velocity, $v_A$. Following our previous works, we bracket this uncertainty by considering two extreme scenarios: one with no reacceleration ($v_A = 0$ km/s) and another with strong reacceleration ($v_A = 40$ km/s), the latter representing an extreme but plausible upper limit on the amount of energy that CRs can gain from interactions with Alfvén waves (see Ref.~\cite{DelaTorreLuque:2023olp} for further details). 
We find that the uncertainty associated with reacceleration can affect our constraints by more than two orders of magnitude. Our benchmark scenario remains close to the most conservative case, corresponding to no reacceleration. However, still the no reacceleration scenario provides constraints that are one order of magnitude weaker than in our benchmark case at masses above a few $\times10^{15}~\mathrm{g}$.
For completeness, Fig.~\ref{fig:fluxBF_vA}, in the Appendix, compares the morphology and spectra of the PBH signal for $M_{\rm PBH}=10^{16}$~g obtained under these three cases: our benchmark model, no without reacceleration, and the case with extreme reacceleration. As can be seen, the different propagation setups produce noticeable variations in both the morphology and the spectra. In particular, the strong reacceleration scenario results in a steeper spectrum and a slightly flatter spatial morphology, as electrons are reaccelerated to higher energies, allowing them to propagate farther from their injection sites before losing their energy.

\begin{figure}[t]
    \centering
    \includegraphics[width=\linewidth]{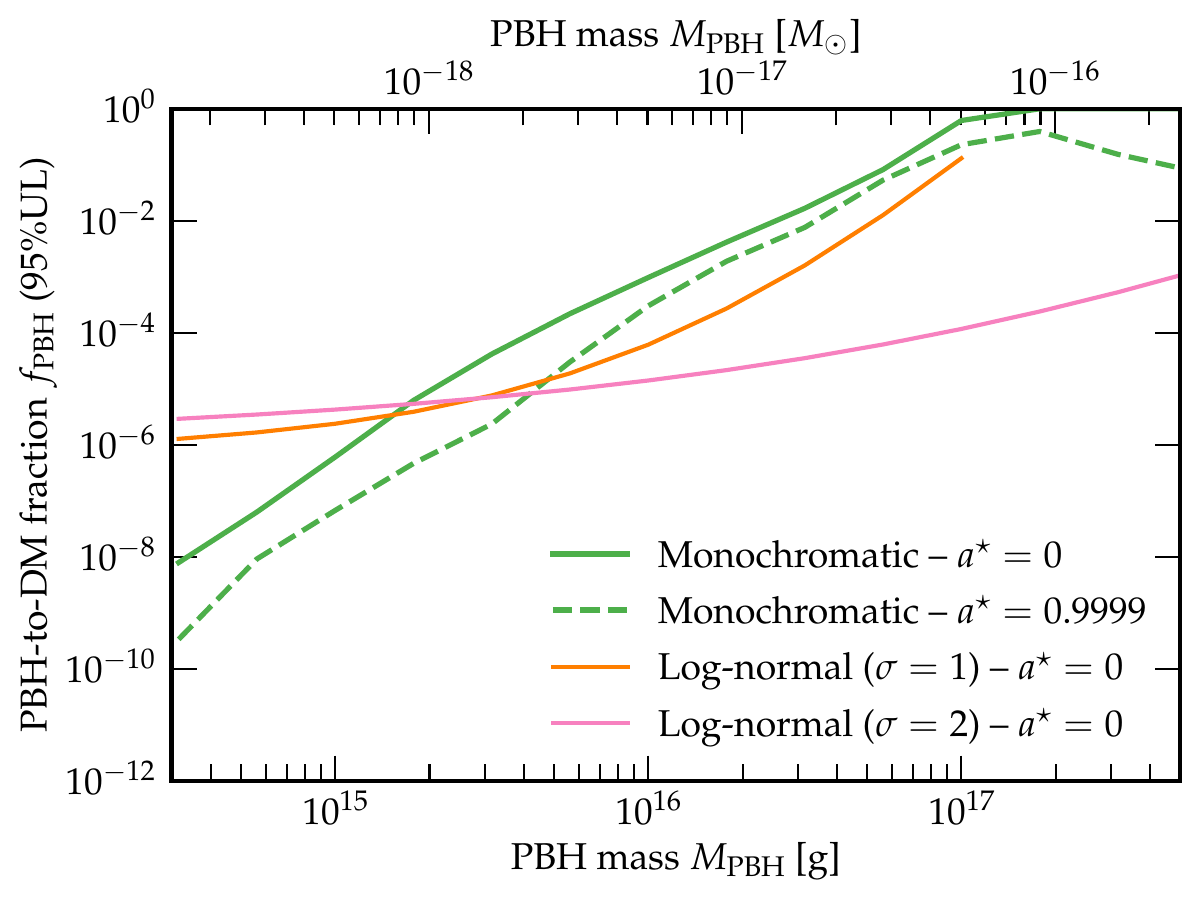}
    \caption{Constraints on the PBH DM fraction for extended log-normal PBH mass functions with $\sigma=1$ (orange) and $\sigma=2$ (pink), and for maximally rotating PBHs with $a_*=0.9999$ (green dashed), compared with the monochromatic Schwarzschild PBH case (solid green).}
    \label{fig:PBH_ext}
\end{figure}

Finally, we compare our benchmark constraints, derived assuming non-rotating PBHs with a monochromatic mass distribution (i.e., a single PBH mass), with those obtained under more realistic assumptions. Specifically, we consider log-normal mass distributions (Eq.~\ref{Eq:lognorm}), which are motivated by several PBH formation scenarios. Since different formation and evolutionary models predict different distribution widths, we adopt two representative values, $\sigma=1$ and $\sigma=2$, commonly used in the literature. We also investigate the impact of PBH spin by comparing our benchmark results with those obtained for a near-maximally rotating population, adopting the theoretical upper limit $a^\star = 0.9999$. 
Fig.~\ref{fig:fluxBF_dist} in the Appendix compares the morphology and spectra of the PBH signal for $M_{\rm PBH}=10^{16}$~g obtained for the benchmark monochromatic Schwarzschild PBH model, extended log-normal mass distributions with $\sigma=1$ and $\sigma=2$, and a near-maximally rotating PBH population. These comparisons illustrate the impact of relaxing the monochromatic, non-rotating benchmark assumption on the predicted spectral and spatial properties of the PBH-induced emission, which lead to a more conservative signal.

\begin{figure}[t]
    \centering
    \includegraphics[width=\linewidth]{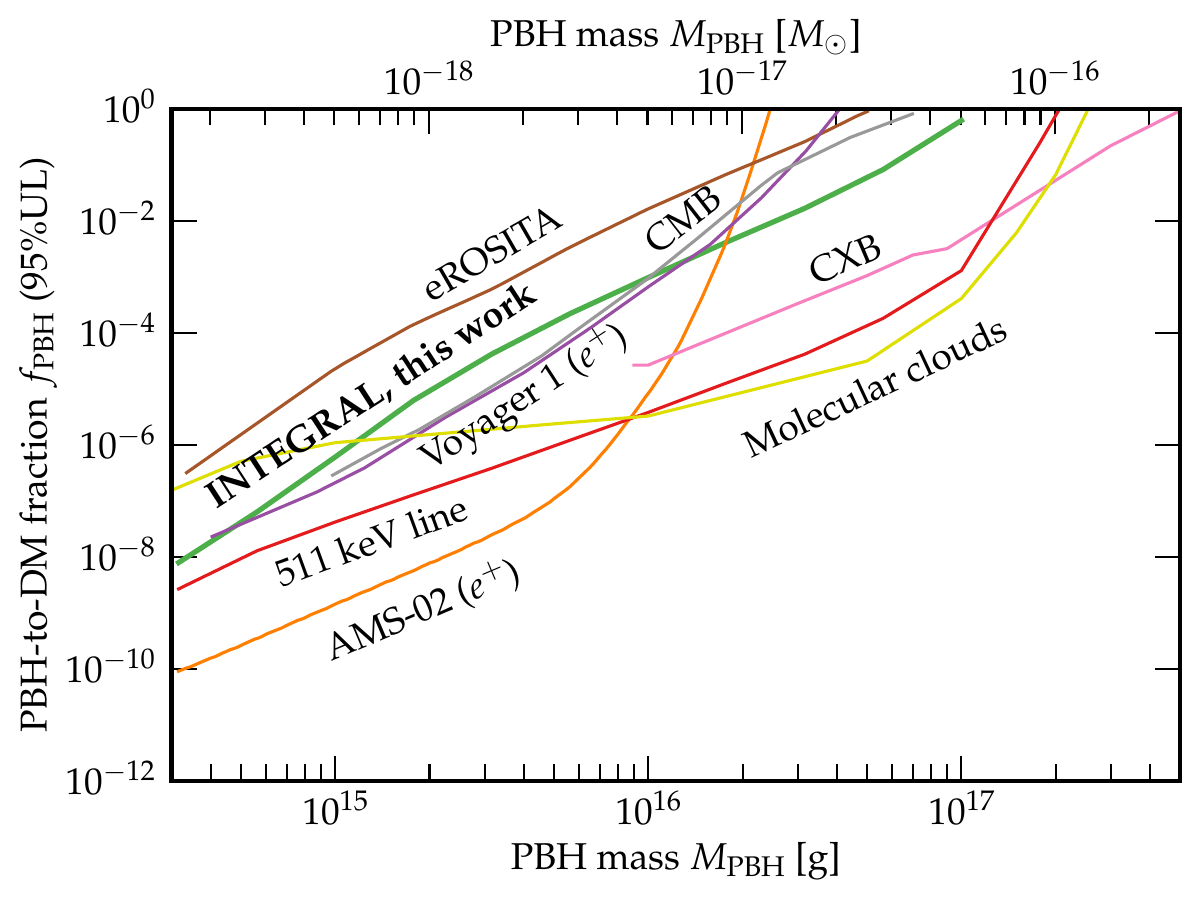}
    \caption{Comparison of our fiducial PBH DM fraction constraints with existing indirect limits from the Galactic 511 keV line morphology~\cite{DelaTorreLuque:2024qms}, the Cosmic X-ray Background (CXB)~\cite{Iguaz2021}, CMB observations~\cite{Clark2017CMB}, eROSITA X-ray constraints~\cite{Balaji:2025afr}, Voyager-1 local electron and positron measurements~\cite{DelaTorreLuque:2024qms}, AMS-02 positron observations~\cite{Huang2025AMS}, and molecular clouds~\cite{Perez:2026cqi}.}
    \label{fig:PBH_litt}
\end{figure}

We show our fiducial limits with the most relevant existing constraints in Fig.~\ref{fig:PBH_litt}. Our results significantly improve upon previous X-ray constraints, surpassing those derived from eROSITA~\cite{Balaji:2025afr} by up to two orders of magnitude in the relevant mass range. They also remain more stringent than the current CMB~\cite{Clark2017CMB} and Voyager-1~\cite{DelaTorreLuque:2024qms, Boudaud2019} constraints above $\sim10^{16}~\mathrm{g}$, remaining very close at lower masses. Notably, the Voyager-1 constraints derived from local electron and positron measurements become significantly weaker above PBH masses of $\sim10^{16}~\mathrm{g}$ in scenarios with low reacceleration, while the IBIS constraints remain stronger in the absence of reacceleration, since they are less affected by CR propagation uncertainties. Nevertheless, the strongest existing constraints are still provided by the morphology of the 511 keV emission line~\cite{DelaTorreLuque:2024qms}, the ionization of molecular clouds close to the GC~\cite{Perez:2026cqi} and the Cosmic X-ray Background~\cite{Iguaz2021}, at high PBH masses, while AMS-02 positron observations~\cite{Huang2025AMS} dominate below $\sim10^{16}~\mathrm{g}$ (with much higher uncertainties associated to reacceleration, though).

%\textcolor{teal}{
%I would show the following plots:
%\begin{itemize}
    %\item Morphology of the DM signals for two/three energy bins comparing different PBH masses, and leaving the CR and NIR background fixed. For a DM fraction that makes the plot visually nice.
    %\item Same for spectrum
    %\item A best-fit plot with combined spectral and morphological information, and a single PBH mass
    %\item Comparison of the morphology and spectra for PBH assuming Monochromatic and non-rotating, monochromatic and Kerr, and one or two mass distribution functions adopted
    %\item Maybe same thing but comparing with reacceleration variations
    %\item Main Constraint plot, with another plot showing how the constraints change with different analyses performed. We have to explain why we vary these analyses based on physical information (e.g. WD mass)
    %\item Constraint compared for the different mass distribution and spin
    %\item constraint vs VA, where we could directly compare with the Voyager constraint for the same choice of VA.
    %\item Maybe an image of the DM signal expected, similar to that in Fig. 9 of https://arxiv.org/abs/astro-ph/0605420 (For the appendix or main text?). Here we could even show the difference of convolving and not convolving with instrument response functions (another plot for the effect in the morphology/spectra of the signals with and w/o these IRFs would be nice). 
%\end{itemize} } 

\section{Discussion and conclusions}
\label{sec:conclusions}

In this work, we have presented a new search for Hawking evaporation signatures from PBHs using 20 years of hard X-ray and soft $\gamma$-ray observations from the IBIS/ISGRI instrument aboard the INTEGRAL satellite. 
By combining the spatial morphology and spectral information of the Galactic diffuse emission within a unified analysis framework, we exploit complementary observables to probe the lower part of the asteroid-mass PBH window, where Hawking evaporation produces energetic electrons and positrons whose IC emission contributes to the X-ray sky. This joint spatial–spectral approach provides additional discriminatory power compared to analyses based only on integrated flux measurements, as it allows the distinct spatial and spectral properties of both the PBH signal and the astrophysical backgrounds to be simultaneously taken into account. By exploiting these complementary features, we can more effectively disentangle potential exotic contributions from uncertainties in the modeling of the diffuse Galactic X-ray emission.
We extract the observations, account for correlations between the morphological and spectral data, and convolve the predicted signals and background components with the instrument response functions.

%The statistical treatment of the IBIS measurements is another important ingredient of our analysis. Since the extracted spatial and spectral measurements are correlated through common INTEGRAL revolutions and shared systematic effects, we reconstruct their covariance matrix using a cluster bootstrap procedure based on the INTEGRAL revolution as the resampling unit. The resulting covariance matrix is regularized through Ledoit--Wolf shrinkage and verified to be numerically stable before being incorporated into the likelihood analysis. This provides a more realistic description of the measurement uncertainties and avoids artificially strong constraints arising from the assumption of independent data points.

A central challenge of this analysis is the accurate characterization of the diffuse X-ray background. We adopt physically motivated models for the dominant astrophysical contributions, namely CR electron-induced IC emission and unresolved accreting WDs, while allowing for additional flexibility through nuisance parameters that account for uncertainties in their normalization and spectral properties. The CR component is allowed to vary through a global normalization factor, whereas the WD contribution is modeled using the spectrum of intermediate polars and connected to the observed Galactic ridge morphology through the near-infrared stellar distribution. This approach combines the predictive power of physically motivated background models with the flexibility required to capture modeling uncertainties, while preserving the spectral and morphological information needed to identify and constrain a possible PBH contribution.

Interestingly, our analyses show that the inclusion of a PBH-induced component can provide a significantly improved description of the IBIS X-ray diffuse observations compared to the background-only hypothesis. The preferred PBH contribution is associated with an additional component with a morphology concentrated towards the Galactic Center, suggesting that the current modeling of the diffuse hard X-ray emission may not capture all relevant emission processes. While such an excess could potentially arise from PBHs or other dark-sector scenarios, an astrophysical origin is probably more likely, given the remaining uncertainties in Galactic X-ray emission. Nevertheless, the preference for an additional component highlights the importance of improving the characterization of diffuse X-ray backgrounds in future studies. Thus, we have adopted a conservative approach and used the data to derive upper limits on the allowed PBH contribution rather than speculating about any significant detection of an exotic component.

We have investigated the robustness of the derived PBH limits against several sources of uncertainty. In particular, we have explored different treatments of the nuisance parameters describing the CR and WD backgrounds, finding that the resulting constraints are relatively stable under reasonable variations of the astrophysical model. We have also quantified the impact of correlations among the IBIS measurements, which produces only moderate changes in the inferred limits but is nevertheless essential for a statistically consistent analysis. Similarly, variations of the Galactic DM density profile have a limited impact, due to the linear dependence of the PBH signal on the density profile and the smoothing effect introduced by the propagation of the injected electrons over Galactic scales.

The dominant theoretical uncertainty arises from the propagation of PBH-produced electrons and positrons, in particular from diffusive reacceleration effects controlled by the Alfvén velocity. By considering extreme propagation scenarios with different levels of reacceleration, we find that the resulting variation in the constraints can reach more than two orders of magnitude. %Our benchmark scenario remains close to the conservative case with negligible reacceleration. 
We further extend our analysis beyond the monochromatic Schwarzschild PBH scenario by considering rotating PBHs and extended log-normal mass distributions, demonstrating the impact of more realistic PBH populations on the expected X-ray signal.

Finally, we compare our constraints with existing bounds from other astrophysical and cosmological observations. The sensitivity achieved with IBIS/ISGRI significantly improves previous X-ray limits in the relevant mass range. Our results also remain competitive with constraints from local measurements of electrons and positrons and remain stronger than CMB constraints at high PBH masses. Although other probes, such as the morphology of the 511 keV emission line, the Cosmic X-ray Background, and AMS-02 positron observations, provide the strongest constraints in the explored mass range, our work demonstrates that hard X-ray observations of the Galactic diffuse emission offer a powerful and largely independent avenue for testing PBHs as a DM candidate.

\vspace{1 cm}
\begin{acknowledgements}
    We thank Roman Krivonos for helpful discussions about the data reduction and the astrophysical models. We also acknowledge Shyam Balaji, Marco Cirelli, and Roman Krivonos for their helpful feedback on the draft. This research is based on observations with INTEGRAL, an ESA project with instruments and science data centre funded by ESA member states (especially the PI countries: Denmark, France, Germany, Italy, Switzerland, Spain), Czech Republic and Poland, and with the participation of Russia and the USA. P.D.L.~has been supported by the Juan de la Cierva JDC2022-048916-I grant, funded by MCIU/AEI/10.13039/501100011033 European Union "NextGenerationEU"/PRTR, and is currently supported by Ramón y Cajal RYC2024-048445-I grant, which is funded by MCIU/AEI/10.13039/501100011033 and FSE+. The work of P.D.L.~is also supported by the grants PID2021-125331NB-I00 and CEX2020-001007-S, both funded by MCIN/AEI/10.13039/501100011033 and by “ERDF A way of making Europe”. P.D.L.~also acknowledges the MultiDark Network, ref. RED2022-134411-T. Some of this work has also benefitted from the support of the European Consortium for Astroparticle Theory in the form of an Exchange Travel Grant to P.D.L.  This project used computing resources from the National Academic Infrastructure for Supercomputing in Sweden (NAISS) under project NAISS NAISS 2024/5-666.
    J.K.~acknowledges support from the research grant {\sl TAsP (Theoretical Astroparticle Physics)} funded by Istituto Nazionale di Fisica Nucleare (INFN), and from the Italian Ministry of University and Research (MUR), PRIN 2022 ``EXSKALIBUR – Euclid-Cross-SKA: Likelihood Inference Building for Universe’s Research'', Grant No. 20222BBYB9, CUP I53D23000610 0006, and from the European Union -- Next Generation EU. J.K.~ also acknowledges support from the Italian Space Agency through the ASI INFN agreement n. 2018-28-HH.0: “Partecipazione italiana al GAPS - General AntiParticle Spectrometer”.
\end{acknowledgements}

\bibliographystyle{apsrev4-1}
\bibliography{paper.bib}

\flushbottom
\newpage
\onecolumngrid
\appendix

\section{Additional figures}

\begin{figure}[t]
    \centering
    \includegraphics[width=\linewidth]{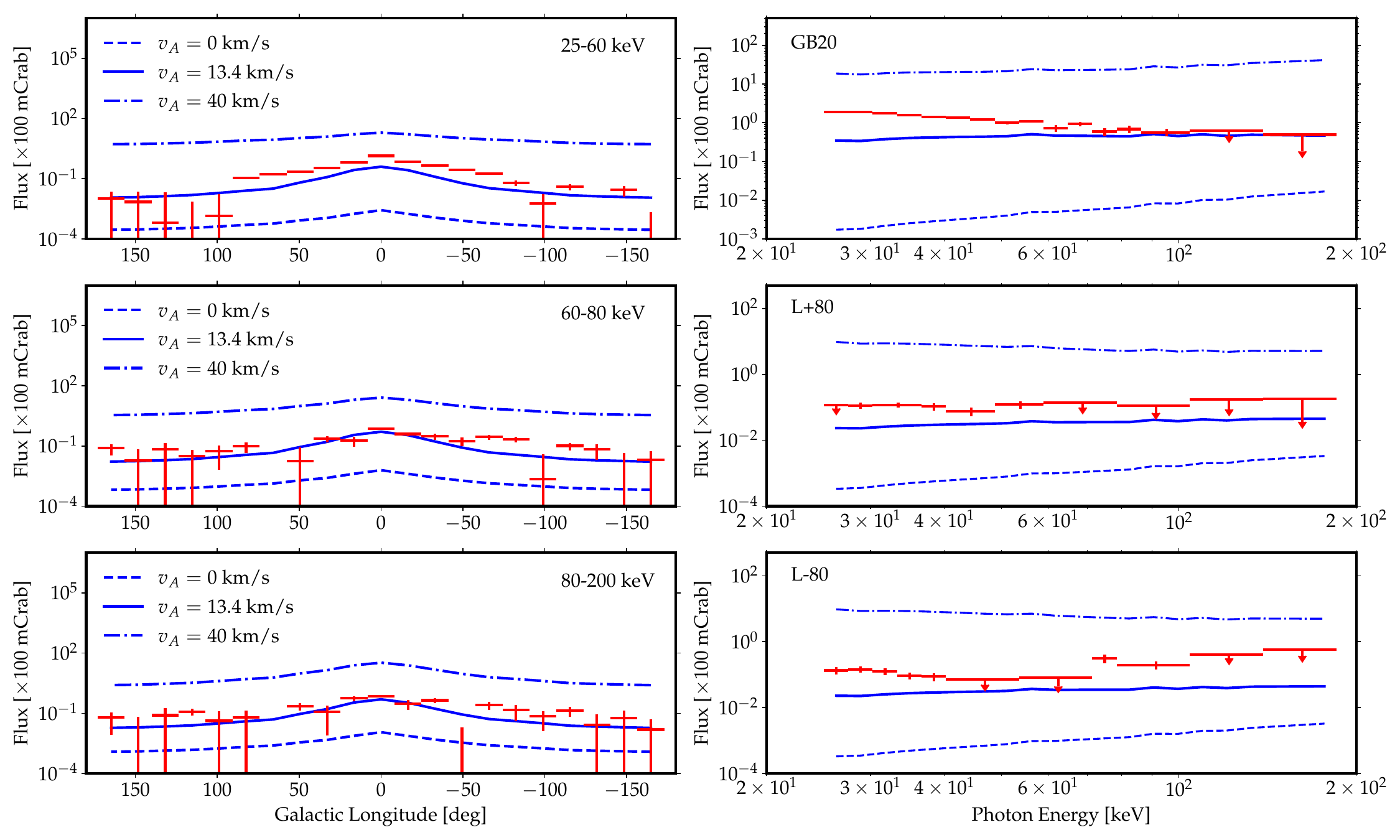}
    \caption{Impact of cosmic-ray reacceleration on the predicted PBH-induced emission for $M_{\rm PBH}=10^{16}$~g. The solid, dashed, and dot-dashed curves show the cases with $v_A=13.4$, $0$, and $40$ km/s, respectively, for a fixed $f_{\rm PBH}=10^{-3}$. Variations in $v_A$ modify both the spectrum and spatial morphology of the IC component, reflecting uncertainties in CR propagation. The flux is displayed on a logarithmic scale.}
    \label{fig:fluxBF_vA}
\end{figure}

\begin{figure}[t]
    \centering
    \includegraphics[width=\linewidth]{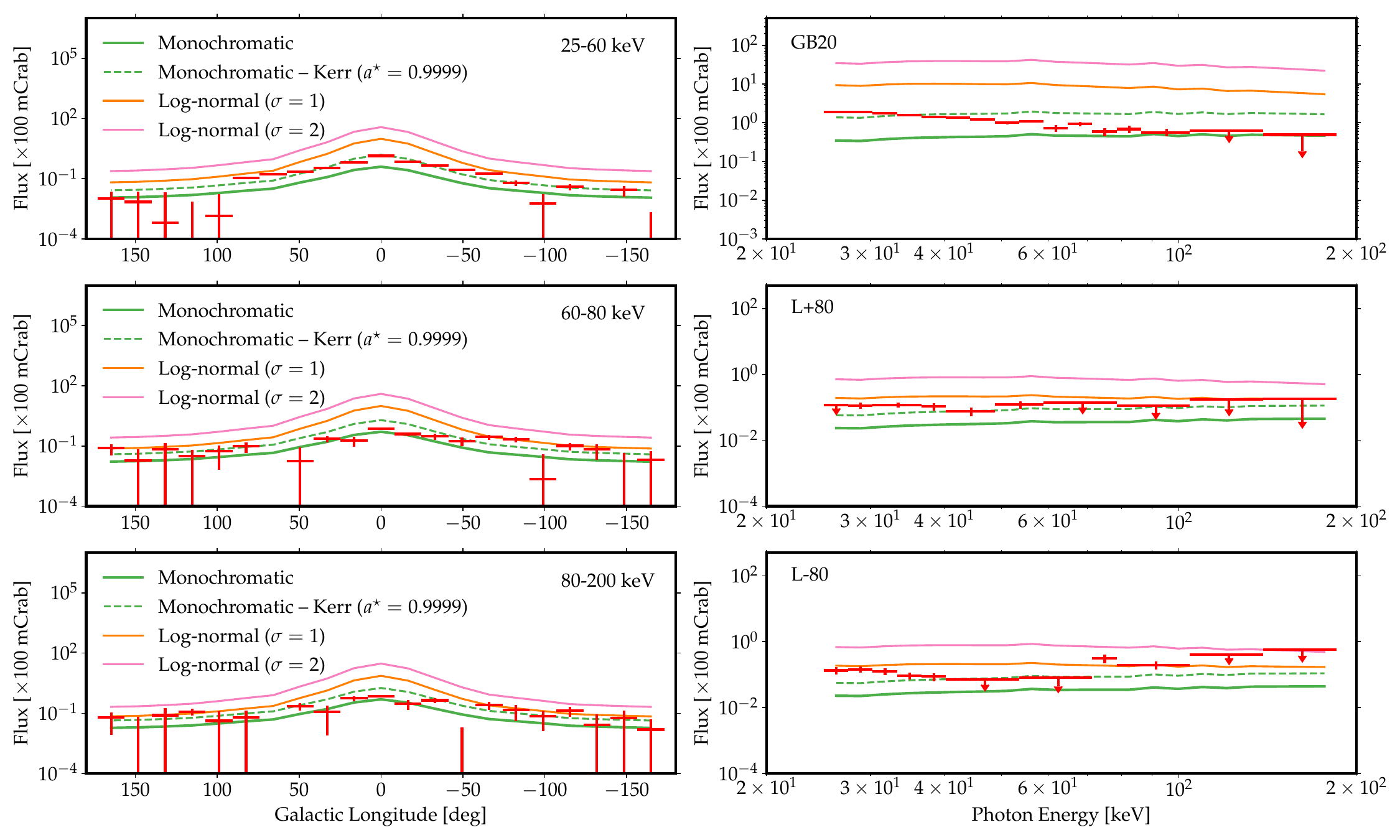}
    \caption{Comparison of the best-fit PBH-induced emission for $M_{\rm PBH}=10^{16}$~g under different assumptions for the PBH population. The solid green curve corresponds to the benchmark monochromatic Schwarzschild case, the dashed green curve to a monochromatic near-extremal Kerr population, and the orange and pink curves to log-normal mass distributions with $\sigma=1$ and $\sigma=2$, respectively. The corresponding abundance is fixed to $f_{\rm PBH}=10^{-3}$. The figure shows the impact of PBH spin and extended mass distributions on both the spectral and spatial morphology of the predicted signal; the flux is displayed on a logarithmic scale to highlight the differences in the spatial distribution.}
    \label{fig:fluxBF_dist}
\end{figure}

In this appendix we show additional figures that illustrate the impact of the main astrophysical and PBH-model assumptions on the predicted PBH-induced X-ray emission, complementing those displayed in the main text. In Fig.~\ref{fig:fluxBF_dist}, we compare the morphology and spectra of the signal for extended PBH mass distributions and rotating PBHs with respect to our benchmark monochromatic Schwarzschild scenario. In Fig.~\ref{fig:fluxBF_vA}, we investigate the effect of CR propagation uncertainties by showing the variations induced by different reacceleration scenarios, parametrized by the Alfv\'en velocity $v_A$.

\end{document}